\documentclass[12pt]{article}
\pdfoutput=1
\usepackage[utf8]{inputenc}
\usepackage{bm}
\usepackage{mathrsfs}
\usepackage{empheq}
\usepackage{verbatim}
\usepackage{bibentry}
\usepackage{hyperref}
\usepackage{slashed}
\usepackage{comment}
\usepackage{subcaption}
\usepackage{multirow}
\usepackage{tikz}
\usepackage{tikz-cd}
\usepackage{color}
\usepackage{tikz}
\usepackage{array}
\usepackage{tablefootnote}
\usepackage{latexsym}
\usepackage{amssymb}
\usepackage{amsmath}
\usepackage{mathtools}
\usepackage{stackrel}
\usepackage{youngtab}
\usepackage{cancel}
\usepackage[nosort]{cite}
\usepackage{graphicx}
\usepackage{epsfig}
\usepackage{tensor}
\usepackage{ulem}

\newcommand{\nin}{\noindent}
\usepackage{tocloft} 
\usepackage{ulem}
\usepackage{soul,xcolor}
\setstcolor{red}

\newcommand{\nn}{\nonumber}
\newcommand{\be}{\begin{equation}}
\newcommand{\ee}{\end{equation}}

\numberwithin{equation}{section}

\usepackage{authblk}

\title{\sf Spinning fields on  S$^d$ and  dS$_d$,  \\ UIRs  and  Ladder operators }
\author[1]
{Vasileios A. Letsios}
\author[2,3] 
{, Mat\'\i as N. Semp\'e$^{2,3}$

and Guillermo A. Silva}
\affil[1] 
{\small \it  Department of Mathematics, King’s College London

Strand, London WC2R 2LS, UK}

\affil[2]
{\small\it Instituto de F\'isica de La Plata - CONICET

Diagonal 113 e/ 63 y 64, 1900 - La Plata, Argentina}
\affil[3]
{\it Departamento de F\'isica, Universidad Nacional de La Plata  

C.C. 67, 1900 - La Plata, Argentina}

\begin{document}
\maketitle
\begin{abstract}
We construct, for spin $0,1,2$ tensor fields on  S$^d$, a set of ladder operators that connect the distinct UIRs of SO$(d+1)$. This is achieved by relying on the conformal Killing vectors of  S$^d$. For the case of spinning fields, the ladder operators generalize previous expressions with a compensating transformation necessary to preserve the transversality condition. We then extend the results to the Exceptional/Discrete UIRs of SO$(d,1)$, again relying on the conformal Killing vectors of de Sitter space. Our construction recovers the conventional conformal primary transformations for the scalar fields when the mass term leads to conformal coupling. A similar approach for the spin-2 field leads to the conformal-like operators found recently.
\end{abstract}

\newpage
\section{Introduction}

The relation between the Unitary Irreducible Representations (UIRs) of the flat space isometry group ISO($d-1,1$) and field equations was elucidated long ago by Eugene Wigner in \cite{Wigner}.  Later on, this relation was extended to maximally symmetric spacetimes in \cite{FR} (see \cite{Di,DN,DW1,Boers,Joung,Anous,Letsios:2023qzq} for related work) . In this work, we will relate different field realizations of SO($d+1$) UIRs on the $d$-dimensional sphere S$^d$ through  differential ladder operators. In addition, we will extend the construction to Exceptional and Discrete UIRs field realizations on de Sitter space (see \cite{zimo,Basile,volo} for nice summaries of SO$(d,1)$ UIRs). As an outcome, we will give a fresh perspective on the conformal-like symmetry operators found in \cite{Letsios:2023tuc} for partially massless gravitons on 4-dimensional de Sitter space. Our setup shows how to generalize these conformal-like symmetry operators to other spacetime dimensions, spins and values of the mass parameter. In particular, for spin-1, we will find a particular tuning of the mass parameter for which the spin-1 field enjoys conformal-like symmetries.

Ladder operators have a long history in physics, their everlasting relevance can be seen for example in recent works where they show up in disparate fields such as black hole Love numbers \cite{Bere} and spinning conformal blocks \cite{SD}. In the present paper, we will generalize previous works \cite{Cardo,Muck} which built first-order differential operators that related solutions of the Klein+Gordon scalar field equation with different masses. We will refer to these differential operators as ladder operators. The main ingredient of the construction required the spacetime to possess a closed conformal Killing vector (CCKV) being also an eigenvector of the Ricci tensor with constant eigenvalue. Here we will focus on \emph{positively curved} maximally symmetric spaces where the eigenvalue condition for CCKV follows naturally from the Einstein character of the space.  We will re-derive previous ladder constructions and extend them to spinning fields.

The paper is structured as follows. In section 2 we start setting the notation and definitions for conformal primaries in conformally flat spaces. In section 3 we discuss and enumerate several properties of conformal Killing vectors (CKV) in maximally symmetric spacetimes relevant for our construction.  In section 4 we construct ladder operators for spin 0,1,2  fields and show how they relate eigenfunctions of the Laplace+Beltrami operator with different eigenvalues. In section 5 we summarize the results and make some proposals for future work. A number of appendices then follow. In Appendix A we spell out the explicit form of Killing and conformal Killing vectors of  S$^3$ and  dS$_4$. In Appendix B we summarize the notation for spherical harmonics on S$^d$.  In Appendix C we spell out the explicit computations for the ladder operators acting on tachyon scalar modes corresponding to  type I Exceptional UIR of SO(d,1). Finally, in  Appendix D we work out the action of ladder operators for vector harmonics on S$^3$.

\section{Primary fields in conformally flat spaces}

In this section, to fix notation, we summarize well known formul\ae~for primary fields in flat space and   define  primary fields on conformally flat spaces.

\subsection{Primary Fields in Flat Space}
Consider $x^\mu$ to be the Cartesian coordinates of $d$-dimensional flat space
$$dx^2=\eta_{\mu \nu}dx^\mu dx^\nu\,.$$
Conformal transformations (CT) are diffeomorphisms  $x'^\mu(x)$ satisfying
\be
\text{\sf Conformal 
Transformation}:~~~~~dx'^2=\Omega^2(x)\,dx^2 .
\ee
Any CT of flat space defines a local scale factor $\Omega(x)$ and a local Lorentz transformation $\Lambda^\mu{}_\nu(x)$ which can be read off from
\be
\frac{\partial x'^\mu}{\partial x^\nu}=\Omega(x)\Lambda^\mu{}_\nu(x)\,.
\label{ct}
\ee
Continuous CT at the infinitesimal level take the form 
$$x'^{\mu}(x) \approx x^\mu+\zeta^\mu(x)+...,~~~|\zeta|\ll1$$ 
with $\zeta^\mu$ satisfying\footnote{A word on terminology: if the factor $\partial\cdot\zeta$ on the rhs of \eqref{CKVe} is zero, we say $\zeta$ is a Killing vector (KV). If the factor is nonzero, we say $\zeta$ is  a proper conformal Killing vector. Occasionally, when the factor is constant, $\zeta$ is called a homothety.}
\be
\text{\sf Conformal Killing vector}:~~~~~\partial_\mu\zeta_\nu+\partial_\nu\zeta_\mu=\tfrac2d(\partial_\rho \zeta^\rho)\eta_{\mu\nu}\,.
\label{CKVe}
\ee
Taking the determinant of \eqref{ct} one finds $$\big(\Omega(x)\big)^d=\det \left|\frac{\partial x'^\mu}{\partial x^\nu}\right|\approx \left|\frac{\partial (x+\zeta)}{\partial x}\right|\approx |\delta^\mu_\nu+\partial_\nu \zeta^\mu|=1+  \partial_\mu \zeta^\mu\,.$$
Thus, 
\be
\Omega(x)\approx1+ \frac1d \partial_\mu \zeta^\mu.
\label{O}
\ee
The   $\zeta$-dependence  of the local Lorentz transformation $\Lambda^\mu{}_\nu(x)$  can be found from
\begin{align}
\Lambda^\mu{}_\nu(x)&=\frac1{\Omega(x)}\frac{\partial x'^\mu}{\partial x^\nu} 
\approx\left(1 - \frac1d \partial_\rho \zeta^\rho\right)(\delta^\mu_\nu+\partial_\nu\zeta^\mu)\nn\\
&\approx\delta^\mu_\nu+\partial_\nu\zeta^\mu-\delta^\mu_\nu\frac1d \partial_\rho \zeta^\rho\,.
\label{llT}
\end{align}
It is useful to define the infinitesimal generators $\sigma_\zeta(x),\omega_\zeta(x)$ associated to  $\zeta^\mu$ as
\be
\sigma_\zeta(x):=\frac1d\partial_\rho \zeta^\rho,~~~(\omega_\zeta)_{\mu\nu}(x):= \frac12( \partial_\nu\zeta_\mu-\partial_\mu\zeta_\nu) \,.
\label{scalefac}
\ee
Then\footnote{Using \eqref{CKVe} we can give the alternative expression 
$$\omega_{\mu\nu}(x)=\frac1d\eta_{\mu\nu}(\partial_\rho \zeta^\rho)-\partial_\mu\zeta_\nu.
$$}
$$ \Lambda(x)=e^{\omega_\zeta(x)}~~~  \text{and}~~~  \Omega(x)=e^{\sigma_\zeta(x)}\,.$$ 

\vspace{2mm}

\nin {\sf Spin zero:} a scalar primary field  with scale dimension $\Delta$ is a field $\Phi$ which under a CT  $x^{\mu} \to x'^{\mu }(x)$ transforms as  
\begin{equation}
\text{\sf Primary scalar field}:~~~~~     \Phi(x)\to \Phi'(x')=\frac{1}{\Omega (x)^\Delta} \Phi(x).
\label{cp}
\end{equation}
To first order in $\zeta$ we have
\begin{align}
    \delta \Phi(x):=&\Phi'(x )-\Phi(x )\nn \\
    =&\Big(1- \frac\Delta d \partial_\mu \zeta^\mu\Big)\underbrace{ (\Phi-\zeta^\mu \partial_\mu \Phi)}_{\Phi(x-\zeta )} -\Phi\nn \\
    &=-\Big(\zeta^\mu \partial_\mu  +\frac\Delta d \partial_\mu \zeta^\mu \Big)\Phi\nn\\
    &=-\left({\cal L}_\zeta +\Delta\, \sigma_\zeta \right)\Phi
    \label{lad1}
\end{align}
The operator inside the parentheses will be extensively used in sections below to construct ladder operators.

\vspace{2mm}

\nin {\sf Spin one}:  under conformal transformations \eqref{ct} a primary vector field $A_\mu$ with scale dimension $\Delta$ transforms as
\begin{equation}
\text{\sf Primary vector field}:~~~~~    A_\mu(x) \to A'_\mu(x')=\frac{1}{\Omega(x)^\Delta} \Lambda_\mu{}^\nu(x)   A_\nu(x).
\label{at}
\end{equation}
The $\Lambda_\mu{}^\nu$ can be found from \eqref{llT} to be
\begin{align}
\Lambda_\mu {}^\nu&\approx  \delta^\nu_\mu+\partial^\nu \zeta_\mu-\frac1d (\partial_\rho \zeta^\rho) \delta_\mu^\nu.
\end{align}
Hence, to first order in $\zeta$ we have
\begin{align}
    \delta A_\mu(x)&=A'_\mu(x)-A_\mu(x)\nn\\
    &=  - \zeta^\rho \partial_\rho A_\mu+
    \partial^\nu \zeta_\mu\,
    A_\nu-\frac1d \partial_\rho \zeta^\rho \,A_\mu- \frac\Delta d \partial_\rho \zeta^\rho A_\mu \nn \\
    &=-\left( \mathcal{L}_\zeta  +  ({\Delta-1 })\, \sigma_\zeta \right) A_\mu,
    \label{1-form}
\end{align}
where ${\cal L}_\zeta$ is the Lie derivative on 1-forms
$$ \mathcal{L}_\zeta  A_\mu=\zeta^\rho\partial_\rho A_\mu+\partial_\mu\zeta^\rho A_\rho\,.$$
To go from the second to the third line in \eqref{1-form} we used \eqref{CKVe}. Equation \eqref{1-form} can also  be written using \eqref{scalefac} as \cite{osborn}
\be
\delta A_\mu=-\zeta^\rho\partial_\rho A_\mu+(\omega_\zeta)_\mu{}^\nu A_\nu- \Delta\, \sigma_\zeta\, A_\mu
\label{Afull}
\ee
Each term in this expression respectively manifest the field, tensorial and scale character of $A_\mu(x)$.

\vspace{2mm}

\nin{\sf Spin two}: a primary symmetric spin-2 field $h_{\mu\nu}$ with scale dimension $\Delta$ transforms as
\begin{equation}
    h_{\mu \nu}\to h'_{\mu \nu}(x')=\frac{1}{\Omega(x)^\Delta} \Lambda_\mu^{\ \alpha}(x)\, \Lambda_\nu^{\ \beta}(x)\, h_{\alpha \beta}(x).
\end{equation}
Proceeding as above one finds 
\begin{align}
    \delta h_{\mu \nu}
    &= -\left(\mathcal{L}_\zeta  +(\Delta-2) \sigma_\zeta\right) h_{\mu \nu}.
\end{align}
For completeness, we quote the Lie derivative on a covariant 2-tensor
\begin{align}
\mathcal{L}_\zeta    h_{\mu \nu} &= \zeta^\rho \partial_\rho h_{\mu \nu}+\partial_\mu \zeta^\rho h_{\rho \nu}+\partial_\nu \zeta^\rho h_{\rho \mu} .
\end{align}

\vspace{2mm}

\nin{\sf Fermions}: a spin-$\tfrac12$ conformal primary $\Psi_a$ with scale dimension $\Delta$  transforms as 
\begin{equation}
    \Psi_a \to \Psi'_a(x')=\frac{1}{\Omega(x)^\Delta} (e^{\frac{1}{2} (\omega_\zeta)^{\mu \nu}  \,\frac{[\gamma_{\mu}, \gamma_\nu]}{4}})^{\ b}_a \,\Psi_b(x)
\end{equation}
with $(\omega_\zeta{})^{\mu \nu}$ defined in  \eqref{scalefac} and $\gamma_\mu$ the Dirac gamma matrices. At the infinitesimal level one obtains
\begin{align}
    \delta \Psi_a&=- \Bigg(\zeta^\rho \partial_\rho-\frac{1}{4}  (\omega_\zeta)^{\mu\nu}\gamma_{\mu\nu}  +\Delta\, \sigma_\zeta\Bigg)  \Psi_a\nn\\
    &=-\big( {\cal L}_\zeta + \Delta \,\sigma_\zeta \big) \Psi_a,
\end{align}
where   $\gamma_{\mu\nu}:=\tfrac12[\gamma_\mu,\gamma_\nu]$ and $\cal L_\zeta$ is the Lie derivative acting on spinors
$${\cal L}_\zeta\psi_{a }= \zeta^\rho \partial_\rho \psi_{a  }  +\frac{1}{4} \partial^\mu \zeta ^\nu (\gamma_{\mu\nu})_a{}^b \psi_{b}.$$

\vspace{2mm}

\nin{\sf Gravitini}: a spin-$\tfrac32$ conformal primary $\psi_{a\mu}$ with scale dimension $\Delta$ transforms under CT as
\begin{equation}
    \psi_{a\mu }  \to  \psi'_{a\mu }(x')= \frac{1}{\Omega(x)^\Delta} (e^{\frac{1}{4} (\omega_\zeta)^{\mu \nu}\, \gamma_{\mu \nu}})_a{}^b \Lambda_\mu{}^{\nu}(x) \psi_{b \nu}(x).
\end{equation}
As expected, the infinitesimal transformation reads
\begin{equation}
    \delta \psi_{a \mu}=-\big( {\cal L}_\zeta + (\Delta-1) \,\sigma_\zeta \big) \psi_{a\mu}
\end{equation}    
with the Lie derivative on gravitini taking the form
\begin{align} 
{\cal L}_\zeta\psi_{a\mu}&= \zeta^\rho \partial_\rho \psi_{a \mu}+ \partial_\mu \zeta^\nu \psi_{a \nu}-\frac{1}{4} \omega_\zeta{}^{\mu\nu}(\gamma_{\mu\nu})_a{}^b \psi_{b \mu}\nn\\
&= \zeta^\rho \partial_\rho \psi_{a \mu}+ \partial_\mu \zeta^\nu \psi_{a \nu}+\frac{1}{4} \partial^\mu \zeta ^\nu (\gamma_{\mu\nu})_a{}^b \psi_{b \mu}.
\end{align}

\subsection{Primary fields in Conformally Flat Spaces}

Conformally flat (CF) spaces have  line element of the form
\be
ds^2=a^2(x)dx^2\,.
\label{CF}
\ee
It is immediate to realize that flat space conformal transformations $x'(x)$ given by \eqref{CKVe} are also CT for conformally flat metrics \eqref{CF}. The resulting conformal factor   can be found simply by noting that under  \eqref{ct} we have
\be
a^2(x')dx'^2=a^2(x')\Omega^2(x)dx^2
=\Upsilon^2(x)a^2(x)dx^2,
\label{cfmet}
\ee
where the conformal factor $\Upsilon(x)$ associated to CT in CF spaces is
\be 
\Upsilon(x):=\frac{a(x')\Omega(x)}{a(x)}\,.
\label{cfac}
\ee
Notice that in CF spaces the conformal factor $\Upsilon(x)$ combines $\Omega(x)$ and $a(x)$.

At the infinitesimal level $\Upsilon(x)$ takes the form
\begin{align}
\Upsilon(x)&\approx \frac{(a(x)+\zeta\cdot\partial a(x)+...)(1+\tfrac1d\partial\cdot\zeta+...)}{a(x)} \nn\\
&\approx1+\frac1d\partial\cdot\zeta+\zeta\cdot\partial\log a+...\nn\\
&\approx1+\frac1d\nabla\cdot\zeta+...
\end{align}
In the last line we used that for conformally flat metrics one has\footnote{The Christoffel symbols for the conformally flat metric \eqref{CF} are $$\Gamma^\mu_{\alpha\beta}=(\delta^\mu_\alpha\delta^\nu_\beta+\delta^\mu_\beta\delta^\nu_\alpha-\eta^{\mu\nu}\eta_{\alpha\beta})\,\partial_\nu(\log a(x))\,. $$ }
$$\nabla\cdot\zeta= \partial_\mu \zeta^\mu+d\, \zeta^\rho\partial_\rho(\log a(x) )\,.$$

 \vspace{2mm}

\nin {\sf Scalars}: a scalar conformal primary in a CF space is  defined as a field $\Phi(x)$ that, under conformal transformations \eqref{ct}, transforms as
\begin{equation}
\text{\sf Conformally flat scalar primary}:~~~~~     \Phi(x)\to \Phi'(x')=\frac{1}{\Upsilon (x)^\Delta} \Phi(x).
\label{cfp}
\end{equation}
At the infinitesimal level, one gets
\begin{align}
    \delta \Phi(x) &=\Phi'(x )-\Phi(x )\nn \\
  &=-\left(\zeta^\mu \partial_\mu  +\frac\Delta d (\nabla_\mu \zeta^\mu) \right)\Phi\,.
    \label{lacf}
\end{align}

\nin{\sf Tensor fields}: a conformally flat spinning primary transforms as\footnote{On a general Lorentz tensor $\Phi_I$, the Lie derivative along a CKV is defined as\cite{ortin}
$${\cal L}_\zeta \Phi_I=\zeta^\rho \partial_\rho\Phi_I-\tfrac{1}{2}  (\omega_\zeta)^{\mu\nu}{\sf D} (S_{\mu\nu})_I{}^J\Phi_J\,.  $$}
$$\text{\sf CF spinning primary}:~~  \Phi_I(x)\to \Phi_I'(x')=\frac{1}{\Upsilon (x)^\Delta} \exp\big(\tfrac12(\omega_\zeta)^{\mu\nu}{\sf D(S_{\mu\nu})}\big)_I{}^J\Phi_J(x).   \,$$
Here ${\sf D} (S_{\mu\nu})_I{}^J$ denotes the particular representation of the Lorentz generators $S_{\mu\nu}$ carried by $\Phi_I$.

In particular, the transformation law for 1-forms is
\begin{equation}
\text{\sf Conformally flat primary vector}:~ A_\mu(x) \to A'_\mu(x')=\frac{1}{\Upsilon(x)^\Delta} \Lambda_\mu{}^\nu(x)   A_\nu(x)
\label{cfA}
\end{equation}
with $\Lambda^\mu{}_\nu(x)$ is defined in \eqref{ct}. At the infinitesimal level it reads (cf. \eqref{Afull})
\begin{align}
\delta A_\mu&=- \zeta^\rho\nabla_\rho A_\mu   -(\nabla_\mu \zeta^\nu)\,A_\nu   - (\Delta-1)\frac{1}{d}\, (\nabla\cdot\zeta)\,A_\mu  \nn\\
&=-\left( \mathcal{L}_\zeta  +  ({\Delta-1 })\, \frac1d\nabla\cdot \zeta \right) A_\mu.
\label{cA}
\end{align}
To obtain the second line we used that the flat CKV fields \eqref{CKVe} are also solutions to \eqref{CKVeqn} for the CF metric \eqref{CF}. 

---

{\nin\footnotesize{\sf Comment on   $\Delta$ and Weyl weight}: in the present context we work with \emph{conformal  isometries} which  lead to a rescalling of the metric by a conformal factor $\Upsilon$, see \eqref{cfmet}. Occasionally, CT are presented as a combination of conformal isometries plus Weyl rescallings that leave the metric invariant \cite{ster}. In this viewpoint, the factors $1/\Upsilon^{\Delta_{\sf Weyl}}$ in the field transformation are understood as arising from the Weyl rescalling. If we adopt this perspective the second term in equation \eqref{cA} corresponds to the Weyl rescalling, hence $\Delta_{\sf Weyl}=\Delta-1$. This result is in agreement with the fact that no Weyl factor is required for gauge fields in $d=4$. }

\section{Conformal Killing Vector identities in  Maximally Symmetric Spaces}

In this section, we introduce the notation to describe the curvature  of maximally symmetric spaces and quote several identities satisfied by Killing and  Conformal Killing Vectors.

---

\nin{\sf  $d$-dimensional maximally symmetric spacetimes (MSS)}: satisfy 
$$  
R_{\mu\nu\alpha\beta}[g] =\frac1{\ell^2}(g_{\mu\alpha}g_{\nu\beta}-g_{\nu\alpha}g_{\mu\beta})$$
\be
R_{\mu\nu}[g] =\frac1{\ell^2}(d-1) g_{\mu\nu}~~~~\text{and}~~~
R[g] =\frac{d(d-1)}{\ell^2}\,.
\label{MSS}
\ee
Here $\ell^2\ne 0$ characterizes the curvature of the space. Positive $\ell^2>0$ correspond to spheres and de Sitter spacetimes in Euclidean and Lorentzian signature respectively, while $\ell^2<0$ give rise to Hyperbolic and anti-de Sitter (AdS) spacetimes.

\vspace{2mm} 

\nin {\sf Killing vectors}:  will be denoted by $k_\mu$
\be
\text{\sf Killing Vector}:~~\nabla_\mu k_\nu+\nabla_\nu k_\mu=0~~\leadsto~~\nabla_\mu k^\mu=0.
\label{kil}
\ee
Contracting this equation with $\nabla^\mu$, commuting covariant derivatives and using Ricci identity one finds 
\begin{equation}
  \text{\sf eqn \eqref{kil}}~\leadsto~  \left\{\begin{array}{c}
\nabla^2  k_\nu= -\frac{d-1}{\ell^2} k_\nu\\
\nabla^\mu k_\mu=0.\, 
\end{array}
\right..
\label{nkil}
\end{equation}
Thus, Killing vectors are transverse eigenvectors of the vector Laplace+ Beltrami operator. It is a well known fact that  in positively curved Euclidean MSS (spheres) the set of Killing vectors have the lowest possible eigenvalues of the vector Laplacian \cite{Higuchi:1986,AGS} (see App.\ref{LadA}).

\vspace{2mm}

\nin {\sf Conformal Killing vectors (CKV)}: they are denoted by $\zeta_\mu$ and satisfy
\be
\text{\sf CKV}:~~~\nabla_\mu \zeta_\nu+\nabla_\nu \zeta_\mu=\frac2d(\nabla\cdot\zeta)\,g_{\mu\nu}.
\label{CKVeqn}
\ee

\vspace{2mm} 

\nin {\sf Proper CKV}: we  denote them by  $c^\mu$. These are CKV with  $\nabla\cdot\bm c\ne0$. The set of CKV-solutions will be  denoted  as $\zeta^\mu=\{k^\mu,c^\mu\}$.

\vspace{2mm}

\nin {\sf Closed CKVs and MSS}: an important result to be extensively used below is that proper CKVs, in {\it curved} maximally symmetric spaces ($\ell^2\ne0$), can be chosen to be  closed\footnote{This fails to be true in flat space as special conformal transformations are not obtained from closed CKVs.}
\be
\text{\sf CCKV}:~~~
\forall c^\mu~\text{in MSS} ~~\leadsto~~
c_\mu=\partial_\mu\Psi.
\label{clo}
\ee
Moreover the potential $\Psi$   is given by its associated scale factor
\begin{equation}
   c_\mu=-\ell^2\, \nabla_\mu \sigma_{\bm c} ~~~~\text{where}~~~ \sigma_{\bm c}:=\frac1d\nabla\cdot \bm c.
    \label{cckv4}
\end{equation}
Notice the need of $\ell^2$ required by dimensional analysis (see \eqref{pS} and \eqref{pD} for the explicit potentials in S$^3$ and dS$_4$).

---

\nin {\footnotesize{\sf Proof}: inserting \eqref{clo} in \eqref{CKVeqn} implies
\begin{equation}
  \nabla_\mu c_\nu=\nabla_\nu c_\mu = \sigma_{\bm c}\, g_{\mu \nu}\,. 
  \label{CCKV}
\end{equation}
Contracting this equation with $\nabla^\nu$ we get
\be
\nabla^\nu  ~\cdot(\nabla_\mu c_\nu   = \sigma_{\bm c}\, g_{\mu \nu})  ~~\leadsto~~  
\nabla^\nu\nabla_\mu c_\nu=\nabla_\mu\sigma_{\bm c}
\label{p1}
\ee
where we used \eqref{cckv4}. Now
\begin{align}
\nabla^\nu\nabla_\mu c_\nu&=\nabla_\mu\nabla^\nu c_\nu+R^\nu{}_{\mu\nu\alpha}c^\alpha\nn\\
&=d\nabla_\mu\sigma_{\bm c}+\frac1{\ell^2}(d \, g_{\mu\alpha}-\delta^\nu_\alpha \,g_{\mu\nu})c^\alpha=d\,\nabla_\mu\sigma_{\bm c}+\frac1{\ell^2}(d-1) c_\mu.
\label{p2}
\end{align}
Since $c_\mu=\partial_\mu\Psi\Rightarrow\nabla^\nu\nabla_\mu c_\nu=\nabla^\nu\nabla_\nu c_\mu$,  comparing \eqref{p1} and \eqref{p2} we obtain \eqref{cckv4}.

}

---

\nin 
Substituting $\nabla_\mu\sigma_c\to c_\mu$ in \eqref{p1} we conclude that 
\begin{equation}
\left\{\begin{array}{c}  \nabla^2 c_\mu= -\frac1{\ell^2}c_\mu\\
   \nabla^\mu c_\mu\ne0.
   \end{array}
   \right.
    \label{cckv3}
\end{equation}
In analogy with \eqref{nkil}, CCKVs are longitudinal eigenvectors of the Laplace +Beltrami operator. In Euclidean MSS (S$^d$), the set of independent CCKVs $\{c_\mu^{(i)}\}$ involves $(d+1)$ solutions which are in one to one correspondence with the lowest longitudinal eigenvector space of the vector Laplacian (cf. \eqref{cckv4} and \eqref{cckv2}) (see \cite{Higuchi:1986}).

For completeness, taking $\nabla^\mu$ in \eqref{cckv4} one obtains 
\begin{equation}
    \nabla^2 \sigma_{\bm c}= -\frac d{\ell^2}\,\sigma_{\bm c}.
    \label{cckv2}
\end{equation}  
This implies that the scale factors $\sigma_{\bm c}$ associated to CCKVs are eigenfunctions of the scalar Laplace+Beltrami operator. In a similar fashion as before, in Euclidean signature, the set of scale factors span the lowest non-trivial  eigenspace of the scalar Laplacian (see also app. B in \cite{AGS})\footnote{Equations \eqref{cckv3}-\eqref{cckv2} are precisely those used in sect. 5.2 of \cite{gizem} to argue that the $\beta_n$-modes with $n=0,\pm1$ correspond to CKV potentials.}.

In the following section, we show that the full spectrum of the scalar Laplacian on spheres S$^d$ can be found from the scale factor of the CCKVs  using ladder operators (cf. \cite{Cardo, Muck}). Our aim in the present paper is to generalize these known results to general tensors on spheres.

---

{\footnotesize \nin {\sf Comment}: it is important to notice, when comparing to the ladder operators to be constructed  below, that CCKVs give no local Lorentz rotation   
\begin{align}
\text{\sf Closed CKV}:~~\partial_{[\mu}\zeta_{\nu]}=0~~&\leadsto~~(\omega_\zeta)^{\mu\nu}=0\nn\\
&\leadsto~~\partial_\mu \zeta_\nu = \frac1d(\partial\cdot \zeta)\, \eta_{\mu \nu}. \nn
\end{align}
Hence, under a CCKV $c^\mu$, a spinning primary  transforms effectively as a scalar
$$\Phi_I\to\Phi'_I(x')=\frac1{\Omega(x)^\Delta}\Phi_I(x). $$
Alternatively, at the infinitesimal level it transforms as
$$\delta \Phi_I= {\mathcal L}_c \Phi_I + (\Delta-s) \sigma_{\bm c} \Phi_I\,.$$

}
 
\section{Laplacian eigenfunctions on positive curvature MSS and Ladder operators}

We now present a systematic procedure to construct all the SO($d+1$)  and SO($d$,1)  UIRs labelled by discrete labels by studying wave equations in positively curved MSS\footnote{For Euclidean signature our procedures leads to all  UIRs of SO$(d+1)$. For the Lorentzian case, we obtain the Exceptional/Discrete series UIRs of SO($d,1$).}. Specifically, we will build  ladder operators ${\cal D},{\cal D}^s$ which, by acting on eigenfunctions of the Laplace+Beltrami operator with a given eigenvalue, raise or lower the eigenvalue. These operators will be constructed out of the CCKVs of the space.

\subsection{Scalars }

\subsubsection{Ladder operators}

We start by assuming that we have an eigenfunction $\Phi$ of the scalar Laplacian 
$$\nabla^2=\tfrac1{\sqrt g}\partial_\mu(\sqrt gg^{\mu\nu}\partial_\nu)$$ 
in a $d$-dimensional maximally symmetric space 
\begin{equation}
    \nabla^2 \Phi =\frac1{\ell^2}\lambda \,\Phi,~~~~\lambda =\Delta(d-1-\Delta)\,.
    \label{eigsc}
\end{equation}
We have parameterized the eigenvalue $\lambda$ with the \emph{scale dimension} $\Delta$. The reason for coining this name will become clear below (see the comment at the end of this subsection)\footnote{See also the comment on $\Delta$ below.}.  It is useful to define the shadow dimension $\Delta^s$ as
\be
\text{\sf Shadow dimension}:~~~~\Delta^s:=d-1-\Delta~~\leadsto~~\lambda=\Delta\, \Delta^s.
\label{shD}
\ee
Since $(\Delta^s)^s=\Delta$, any given Laplace+Beltrami eigenfunction $\Phi$ can be characterized as having either $\Delta$ or $\Delta^s$ scale dimension. 

To make contact with known expressions, for positive curvature maximally symmetric spaces ($\ell^2>0$) we have:
\begin{itemize}
\item {\sf Euclidean signature}: scalar spherical harmonics $\Phi_k$ in  S$^d$ correspond to UIRs of SO$(d+1)$\footnote{For ease of notation in this paragraph we denoted with $k$ the first of the quantum numbers in ${\bm l}=(k,l_{d-1},...l_1)$. This is the only one relevant for the discussion. See  app. \ref{SphArm} for notation.}. From 
\be
\nabla^2\Phi_k=-\frac{k(k+d-1)}{\ell^2}\Phi_k~~~~~k=0,1,2,...,
\label{SH}
\ee
we see they  map  to the $\Delta=-k$ solutions in \eqref{eigsc}. 
\item{\sf Lorentzian signature}: tachyonic scalars in   dS$_d$ with
\be
\ell^2m^2= -n(n+d-1),~~~n=0,1,2,...
\label{mass}
\ee
correspond to type I Exceptional UIRs of SO$(1,d)$ \cite{Boers,zimo,gizem,EM}. From 
\be 
(\nabla^2-m^2)\Phi_n=0,
\label{tachy}
\ee
we conclude that they correspond to  $\Delta=-n$ in \eqref{eigsc}. 

Intriguingly \eqref{tachy}  possesses a (shift) invariance under \cite{hinter}\footnote{In  Euclidean signature, the $\lambda_n$-modes  correspond to the zero modes of ${\cal O}= \ell^2\nabla^2+n(n+d-1)$ on S$^d$.}
\be
\phi\mapsto\phi+\lambda_n,~~~~\lambda_n=s_{I_1...I_n}X^{I_1}...X^{I_n},
\label{gm}
\ee
where $s_{I_1...I_n}$ is a real traceless symmetric constant tensor and $X^I$ are the coordinates of the standard ambient space realization of de Sitter space \cite{StrVol,AnMus}. As stressed in \cite{fola} and discussed in app. \ref{SSs}, this invariance should be understood as gauged. For $n = 0$, \eqref{gm} becomes the familiar constant shift symmetry of the free, massless scalar (see also \cite{law}). The gauging of the shift symmetry for massless scalars in dS$_2$  was performed recently in \cite{gizem}.

In the following sections, as well as in appendices \ref{SSs} and \ref{LaddDS}, we discuss the consequences of the set of zero mode solutions arising from the Wick rotation of the \emph{irregular} solutions on the sphere (Legendre $Q$-function).
\end{itemize}
Since two scale dimensions $(\Delta,\Delta^s)$ are associated to each  $\Phi$ satisfying \eqref{eigsc}, we define operators ${\cal D},{\cal D} ^s$, built out of  CCKVs  $c^\mu$,   as
\begin{align}
     {\cal D} \Phi &:= c^\mu \nabla_\mu \Phi +\Delta \,\sigma_{\bm c}\, \Phi, \nn\\ 
     {\cal D}^s  \Phi &:= c^\mu \nabla_\mu \Phi +\Delta^s \,\sigma_{\bm c}\, \Phi  .
    \label{fipp}
\end{align}
Notice these expressions resemble the infinitesimal form of conformal scalar primary transformations \eqref{lacf}.

In the following, we  show that the action of ${\cal D,D}^s$
on modes $\Phi$ satisfying \eqref{eigsc}, i.e. with scale dimension $\Delta$, generates 
eigenmodes $\Phi'$ and $\Phi''$,  
$$\Phi':={\cal D} \Phi,~~~~\Phi'':= {\cal D}^s  \Phi,$$ 
which are eigenfunctions of the Laplace+Beltrami operator with shifted scale dimension. Explicitly, 
\begin{align}
\nabla^2\Phi' =&\frac{1}{\ell^2}\Delta'(d-1-\Delta')\,\Phi'~~~\text{with}~~\Delta'=\Delta+1,  \nn\\
\nabla^2\Phi''=&\frac{1}{\ell^2}\Delta''(d-1-\Delta'')\,\Phi''~~~\text{with}~~\Delta''=\Delta-1.
\label{shiftSc}
\end{align}
Alternatively, ${\cal D},{\cal D}^s$  can be viewed as shifting the shadow dimension as  $\Delta^s\mapsto\Delta^s\mp1$ 
$$\nabla^2\Phi=\frac{1}{\ell^2}\Delta\Delta^s\,\Phi~~\Rightarrow~~
\left\{\begin{array}{ll}
\nabla^2\Phi'=\frac{1}{\ell^2}\Delta'\,\Delta'^s\,\Phi',&~ \Delta'^s=\Delta^s-1,\\
\nabla^2\Phi''=\frac{1}{\ell^2} \Delta'' \Delta''^s \,\Phi'',&~\Delta''^s=\Delta^s+1.
\end{array}
\right.$$
This means that ${\cal D},{\cal D}^s$ raise and lower the associated scale dimensions of the Laplace eigenfunction $\Phi$. The commutation relations with the Laplace+Beltrami operator are
\begin{align}
    [{\ell^2}\nabla^2, {\cal{D}}]\Phi &= (   d-2-2\Delta )\,{\cal{D}}\Phi, \nn\\
    [{\ell^2}\nabla^2, {\cal{D}}^s]\Phi &=  (d-2-2\Delta^s)\,{\cal{D}}^s\Phi.
    \label{DA}
\end{align}

---

\nin {\footnotesize{\bf Proof}: we assume we are in a MSS, then
\begin{equation}
    [{\cal D} ,\nabla^2] \Phi =-(\nabla^2 c^\mu) \nabla_\mu \Phi -2 \sigma_{\bm c}\, \nabla^2\Phi -\frac{d-1}{\ell^2}c^\mu \nabla_\mu \Phi-\Delta (\nabla^2\sigma_{\bm c}) \Phi-2 \Delta\, (\nabla^\rho \sigma_{\bm c}) \nabla_\rho \Phi.
\end{equation}
Using  \ref{cckv3},\ref{cckv2}  we get
\begin{align}
    [{\cal D},\ell^2\nabla^2] \Phi&= c^\mu \nabla_\mu \Phi-2 \sigma_{\bm c} \Delta\,\Delta^s \Phi-(d-1)c^\mu \nabla_\mu \Phi+d\, \sigma_{\bm c} \,\Delta \, \Phi+2 \Delta\, c^\mu \nabla_\mu \Phi \nn\\
    &= (2 \Delta-d+2)\,c^\mu \nabla_\mu \Phi+ \sigma_{\bm c} \,\Delta\,(-2  \Delta^s+d   ) \Phi\nn\\
    &=(d-2\Delta^s) [c^\mu \nabla_\mu \Phi+\Delta\, \sigma_{\bm c}  \, \Phi]\nn\\
    &=(d-2\Delta^s)\,{\cal D}\Phi.
    \label{LadScal}
\end{align}
We conclude that 
\be
[{\cal D} , \nabla^2]=\frac1{\ell^2}(d-2\Delta^s)\, {\cal D},~~~\qquad ~~[{\cal D}^s, \nabla^2]=\frac1{\ell^2}(d-2\Delta)\, {\cal D}^s\,. 
\label{t}
\ee
The second identity follows from $(\Delta^s)^s=\Delta$. Therefore, it now follows from \eqref{t} that
\begin{align}
\nabla^2\Phi'=\nabla^2 ({\cal D}\Phi)&={\cal D}\nabla^2\Phi-\frac1{\ell^2}(d-2\Delta^s)\, {\cal D}  \Phi \nn\\
&= \frac1{\ell^2}(\Delta\Delta^s-d+2\Delta^s)\, {\cal D}  \Phi\nn\\
&= \frac1{\ell^2}(\Delta+1) (d-1-(\Delta+1)) \,\Phi'\nn\\
&= \frac1{\ell^2}(\Delta+1) (\Delta+1)^s \,\Phi' .
\label{}
\end{align}
Similarly, we have
\begin{align}
\nabla^2\Phi''&= \frac1{\ell^2}(\Delta^s+1) (\Delta^s+1)^s \,  \Phi''\nn\\
&=\frac1{\ell^2}(\Delta-1)(d-\Delta)\Phi''  .
\label{Ds}
\end{align}

---

}

\nin   {\sf Comment on conformal coupling}:  the rhs in \eqref{LadScal} vanishes for $\Delta^s=\frac{d}{2}$ which is equivalent to $\Delta=\frac{d}{2}-1$. This means that the $\cal D$-operators preserve the $\lambda = (\frac{d}{2}-1)\frac{d}{2}$ eigenspace. Notice that the value $\Delta=\frac{d}{2}-1$ coincides with the \emph{classical scale dimension} of a scalar field in $d$-dimensions. This situation has a neat interpretation  as $\lambda=\ell^2m^2= (\frac{d}{2}-1)\frac{d}{2}$ is precisely the   `mass' value equivalent to conformal coupling to the scalar curvature that guarantees Weyl invariance of the field equation \cite{BD} and therefore invariance under the CKVs of the MSS (cf. right column top in fig.\ref{Delta diagram}).

\subsubsection{Building discretely labelled UIRs  on $\ell^2>0$ MSS}

{\sf Sphere}: the constant function on the sphere $\Phi_0(x)=const.$  has vanishing Laplace eigenvalue, $\lambda=0$.  The  scale dimensions associated to $\Phi_0$ are $\Delta=0$ and  $\Delta^s=d-1$. From \eqref{fipp} we learn that ${\cal D}\Phi_0=0$ vanishes identically. However, the action of ${\cal D}^s$ for any of the $(d+1)$ CCKVs $c^\mu$ gives rise  $(d+1)$  non-trivial eigenfunctions\footnote{The $(d+1)$ CCKVs are customarily denoted as $\bm D,\bm K_i$ with $i=1,...,d$.}
$${\cal D}^s\Phi_0 \propto\sigma_{\bm c}.$$
Comparing with \eqref{cckv2} and in accordance with \eqref{Ds},  we recognize that the set $\{\sigma_{\bm c}\}$ built from each of the  $(d+1)$ linearly independent $c^\mu$'s,  coincides with the first non-trivial harmonics on S$^d$, i.e. those corresponding to $\lambda=-d$ or $k=1$ in \eqref{SH}. Naturally, they mix under the action of the Killing vectors. If we continue recursively, acting with $({\cal D}^s)^k$ on $\Phi_0 $,  we obtain scalar spherical harmonics with all possible eigenvalues of the Laplacian (see fig.\ref{SpH}). The remaining states on the multiplet can then be found by Lie derivative actions along the Killing vectors on the sphere. The standard inner product on S$^d$ gives rise to the ${\bm l}=(k,0,...0)$ finite dimensional UIR of SO($d+1$).

\vspace{2mm}

\begin{figure}[h]
\centering
\captionsetup{width=.85\linewidth}
\begin{tikzcd}  
 \vdots & &   \vdots  \arrow[bend right=-30]{d}{{\cal D}  }   &  \vdots \\ 
-\nabla^2=2(d+1)    & &   ({\cal D}^s )^2\Phi_0 \arrow[bend right=-30]{u}{{\cal D}^s }  \arrow[bend right=-30]{d}{{\cal D}\lvert_{_{\Delta=2}} }   &  \Delta=-2\\   
 -\nabla^2=d    & &  {\cal D}^s\Phi_0\propto\sigma_{\bm c}(x)   \arrow[bend right=-30]{d}{{\cal D }\lvert_{_{\Delta=1}}} \arrow[bend right=-30]{u}{{\cal D}^s\lvert_{_{\Delta^s=d}}}   &  \Delta=-1\\
-\nabla^2=0  &    & \Phi_0(x)=cte. \arrow[bend right=-30]{d}{{\cal D}\lvert_{_{\Delta=0}}={\cal L}_c} \arrow[bend right=-30]{u}{{\cal D}^s\lvert_{_{\Delta^s=d-1}}}& \Delta=0 \\
  &   &  0 & 
\end{tikzcd}
\caption{Construction of the full tower of scalar spherical harmonics on spheres out of $\Phi_0(x)=const.$. The action of ladder operators ${\cal D}^s$, built out of CCKVs $c^\mu$, takes us up indefinitely along the tower. At each level, denoted by $\Delta=-k$, the remaining states of the UIR multiplet are obtained by acting 
with Lie derivatives along the Killing vectors of the sphere.}
\label{SpH}
\end{figure}

\nin {\sf de Sitter space and tachyonic shift symmetries}: the Lorentzian signature case presents additional features. We start by reminding the reader that from
$$ds^2=\ell^2(d\theta^2+\sin^2\theta d\Omega^2_{d-1}),$$
with $d\Omega^2$ the round sphere metric, we obtain the de Sitter metric by making the Wick rotation 
\be
\theta\mapsto\frac\pi2-it.
\label{wick}
\ee
The result is
\be
ds^2=\ell^2(-dt^2+\cosh^2t\, d\Omega^2_{d-1}).
\label{glob}
\ee

It is straightforward to show that a constant function on de Sitter spacetime,  which we denote $\Phi_0(x)=const.$, is a seed for all the gauge modes $\lambda_n$ displayed in \eqref{gm} (see left column in fig.\ref{Delta diagram}). As well known, the gauge modes can be written as homogeneous polynomials in embedding space coordinates. Their explicit form in de Sitter global coordinates \eqref{glob} is obtained by performing the Wick rotation \eqref{wick} on the standard spherical harmonics ${\sf Y}_{\bm l}(\theta,\bm \Omega)$ (see app. \ref{SSs})\footnote{We succinctly denote by $\bm l$ the full set of quantum numbers labeling the harmonic (see app. \ref{SphArm}).}.
 
\begin{figure} 
\captionsetup{width=.85\linewidth}
\begin{tikzcd}  
m^2=\frac{d}{2}(\frac{d}{2}-1)  & & \overbrace{{\cal D}...{\cal D}}^{\frac d2-1~\text{times}} \!\!\!\tilde\Phi_0(x)~~~~\arrow[bend right=-30]{d}{{\cal D}^s }   \arrow[out=105,in=65,loop,looseness=3]{}{{\cal D}\lvert_{_{\Delta_c=\frac d2-1}}}  &  \Delta=\frac{d}{2}-1\\   
  \vdots   & & \arrow[bend right=-30]{u}{\cal D} \arrow[bend right=-30]{d}{{\cal D}^s\lvert_{_{\Delta^s=d-3}} } \vdots    &  \vdots\\   
m^2=d-2    &0 & {\cal D}\tilde\Phi_0(x) \arrow[bend right=-30]{d}{{\cal D}^s\lvert_{_{\Delta^s=d-2}} } \arrow[bend right=-30]{u}  {\cal D}    &  \Delta=1\\
   m^2=0  & \Phi_0(x)=cte  \arrow[bend right=-30]{d}{{\cal D}^s\lvert_{_{\Delta^s=d-1}} } \arrow[bend right=-30]{u}{{\cal D}\lvert_{_{\Delta=0}}={\cal L}_c }
   & \tilde\Phi_0(x) 
   \arrow[bend right=-30]{d}{{\cal D}^s\lvert_{_{\Delta^s=d-1}}} 
   \arrow[bend right=-30]{u}{{\cal D}\lvert_{_{\Delta=0}}  ={\cal L}_c}& \Delta=0\\
m^2=-d  &{\cal D}^s \Phi_0\propto\sigma_c  \arrow[bend right=-30]{d}{{\cal D}^s\lvert_{_{\Delta^s=d}} }  \arrow[bend right=-30]{u}{{\cal D}\lvert_{_{\Delta=-1}} }    & {\cal D}^s\tilde{\Phi}_0 (x) \arrow[bend right=-30]{d}{{\cal D}^s\lvert_{_{\Delta^s=d}} }  \arrow[bend right=-30]{u}{{\cal D}\lvert_{_{\Delta=-1}}  } 
&\Delta=-1\\
 m^2=-2 (d+1)  & {\cal D}^s\sigma_c  \arrow[bend right=-30]{d}{{\cal D}^s } \arrow[bend right=-30]{u}{{\cal D}\lvert_{_{\Delta=-2}} }   & 
 {\cal D}^s{\cal D}^s\tilde{\Phi}_0 (x) \arrow[bend right=-30]{d}{{\cal D}^s } \arrow[bend right=-30]{u}{{\cal D}\lvert_{_{\Delta=-2}}  }
& \Delta=-2  \\
\vdots  &  \arrow[bend right=-30]{u}{\cal D }  \vdots   & \arrow[bend right=-30]{u}{{\cal D} }\vdots   & \vdots 
\end{tikzcd}
\caption{On the left column we show that all regular gauge modes \eqref{gm} at different levels are connected by the ${\cal D, D}^s$ ladder operators acting on the constant function ${\Phi}_0$. On the right column we display that the irregular modes are also connected through ${\cal D, D}^s$ ladders. Amusingly in even dimensions the action of ${\cal D}$-ladder stops at level $\Delta=\frac d2-1$ (cf. top right column).}
\label{Delta diagram}
\end{figure}

However, in Lorentzian signature at level $\Delta=0$, a second (non-trivial) solution exists which we denote as  $\tilde\Phi_0(x)$. Explicitly, $\Phi_0$ and ${\tilde\Phi}_0$ are solutions to the massless KG equation independent of the (spatial) coordinates on the sphere in \eqref{glob}.  They are obtained as
\be
\partial_t\left((\cosh t)^{d-1}\partial_t\Phi\right)=0~~\leadsto~~\Phi=c_1 +c_2\, \int^t\frac1{(\cosh t')^{d-1}}dt' .
\label{zMode}
\ee
The term involving the integral in \eqref{zMode} is what we call ${\tilde\Phi}_0$. It can be understood as the Lorentzian version of the Legendre $Q$-function (see App. \ref{SSs}). Interestingly,  both ${\cal D}$ and ${\cal D}^s$ act non-trivially on it\footnote{The $t$-dependent \emph{zero mode} for the massless scalar in $4d$ de Sitter was originally discussed in \cite{Allen} (see also \cite{AF,GK,GT,TT}).}. Amusingly, in even dimensions, the action of $\cal D$ stops at level $\Delta=\frac{d}{2}-1$. At this particular value the  mass in the KG equation coincides with conformal coupling and ${\cal D}$ commutes with the Laplace operator (cf. \eqref{t}). Equivalently, for $\Delta=\frac{d}{2}-1$ the ladder ${\cal D}$ preserves the space of solutions of the Klein+Gordon equation. $\cal D$ is nothing but the infinitesimal form of a combined Weyl and conformal diffeomorphism transformation of the scalar field (see comment below \eqref{cA}).

\subsubsection{Action of ${\cal D,D}^s$  on dS$_d$ tachyonic mode functions}
\label{LaddDS}

Although the construction works for any CCKV in positive curvature MSS, simple  expressions are found for:
\begin{align}
&\text{\sf Sphere}:\qquad~~~~\bm c=\sin\theta\,\bm\partial_\theta&\qquad&\text{ in } \quad \frac{ds^2}{\ell^2}=d\theta^2+\sin^2\theta\, d\Omega_{d-1}^2 \nn \\
&\text{\sf de Sitter}:\qquad~~\bm c=\cosh t\,\bm\partial_t&\qquad&\text{ in }\quad \frac{ds^2}{\ell^2}=-dt^2+\cosh^2\!t\, d\Omega_{d-1}^2\,. \nn
\end{align}
Our construction shows that the set of states  $\{{\cal D}^s...{\cal D}^s \Phi_0\}$ on the left column in fig. \ref{Delta diagram} correspond to the regular gauge modes discussed in the context of scalar tachyons in de Sitter (see \cite{BEM},\cite{gizem},\cite{zimo}).

In this section we spell the details of action of ${\cal D,D}^s$ on the mode functions for scalar tachyons satisfying
$$(-\nabla^2+m^2)\Psi=0$$
with  $m^2=-n(n+d-1)$. The mode solutions to the Klein+Gordon equation in global coordinates  take the form (see app. \ref{SSs} for notation and  details)
$$\Psi_{\bm l}(t, \bm{\Omega} ) =   f_l({t}) \, {\sf Y}_{\bm l}(\bm{\Omega})$$
where
$$f_l(t)\in \left\{\left( \cosh{t}   \right)^{-(d-2)/2}P^{-\left( l  +\frac{d-2}{2}   \right)}_{n+\frac{d-2}{2}}(i \sinh{t}),\,\left( \cosh{t}   \right)^{-(d-2)/2}Q^{  l  +\frac{d-2}{2}}_{n+\frac{d-2}{2}}(i \sinh{t})\right\}.$$
The relevant feature of the following discussion will be on the shift of the mass $m^2$, parametrized by $\Delta=-n$. Therefore, in the present section, we suppress the angular quantum numbers $\bm l$ and simply denote the tachyon field with a subscript $n$. We write
\begin{align}
\Phi_n(t,\bm\Omega)&=\left( \cosh{t}   \right)^{-(d-2)/2}P^{-\left( l  +\frac{d-2}{2}   \right)}_{n+\frac{d-2}{2}}(i \sinh{t})\, {\sf Y}_{\bm l}(\bm{\Omega})
\label{Pmod}\\
\tilde\Phi_n(t,\bm\Omega)&=\left( \cosh{t}   \right)^{-(d-2)/2}Q^{-\left( l  +\frac{d-2}{2}   \right)}_{n+\frac{d-2}{2}}(i \sinh{t})\, {\sf Y}_{\bm l}(\bm{\Omega})\,.
\label{Qmod}
\end{align}

It is sufficient to compute the action of the ladder ${\cal D}^s \Phi_n$ with respect to only one (out of $d$) CKV $c^{\mu}$. The ladders with respect to the remaining CKV can be found using the de Sitter isometries $k_\mu$ as
\begin{align}
[{\cal L}_{\bm k} ,  {\cal D}_{\bm c}^s ] \Phi_n &= \left({\cal L}_{[\bm k,\bm c]}+(n+d-1) \,\sigma_{\bm [\bm k,\bm c]}\,  \right) \Phi_n\nn\\
&={\cal D}_{\bm [\bm k,\bm c]}^s \Phi_n.
\end{align}
In this equation, we have explicitly denoted the CKV dependence of the ladder operator by a subscript,  ${\cal D}^s_{\bm c}$.  

We start by choosing, in global coordinates, the CKV  given by
\begin{align}
    c_{\mu} =- \partial_{\mu}  \sinh{t} ~~\leadsto~~\sigma_{\bm c} =  \sinh{t}.
\end{align}
This choice is particularly convenient since it does not involve the  spatial coordinates $\bm \Omega$.

\vspace{2mm}

\nin {\sf Rising ladder}: our aim is to compute 
\begin{align}
     {\cal D}^s \Phi_n = \left({\cal L}_{\bm c}+(n+d-1) \,\sigma_{\bm c}\,  \right) \Phi_n  .
\end{align}
More explicitly, one has
\begin{align}
     {\cal D}^s \Phi_n =\left(\cosh{t} \, \partial_t+(n+d-1) \,\sinh{t}\,  \right) \Phi_n  .
\end{align}
Inserting \eqref{Pmod} into this equation, and introducing  $z= i \sinh{t}$, we straightforwardly find
\begin{align}
     {\cal D}^s \Phi_n &= -i\,{\sf Y}_{\bm l}(\bm \Omega)\,  (\cosh{t})^{-\frac{d-2}{2}} \nonumber \\
     &\times \left( (z^{2}   -1)\frac{\partial}{\partial z}+\left(n+\frac{d}{2}\right)z   \right) P^{-\left( l  +\frac{d-2}{2}   \right)}_{n+\frac{d-2}{2}}(z)\, .
\end{align}
Using (eqn. 14.10.4 in \cite{dlmf})
$$\left(  (x^{2}-1)  \frac{  \partial  }{\partial x} + (\nu +1)x  \right)P^{-\mu}_{\nu}(x)  =  (\nu + \mu +1)P^{-\mu}_{\nu+1}(x)$$
one obtains
\begin{align}
     {\cal D}^s \Phi_n = - i (n+l + d-1) \Phi_{n+1}.
\end{align}
Similarly, for the $Q$-modes \eqref{Qmod} we obtain  
\begin{equation}
    {\cal D}^s \tilde\Phi_n=-i(1+n-l) \tilde \Phi_{n+1} .
\end{equation} 
Some comments are in order: 

\vspace{1mm}

\nin  i. Gauge modes $l\le n$ of $\Delta=-n$ are mapped onto gauge modes of the $\Delta = -(n+1)$ theory. 

\vspace{1mm}

\nin  ii. The lowest physical modes,  i.e.  $l=n+1$, of the $\Delta = -n$  theory are mapped onto  gauge modes of the $\Delta = -(n+1)$ theory. 

\vspace{1mm}

\nin iii. All higher physical modes of the $\Delta = -n$  theory (i.e.  modes with $l   \geq n+2$) are mapped onto physical modes of the $\Delta = -(n+1)$ theory. 

\vspace{1mm}

\nin iv.  $Q$-modes with $l=n+1$ (physical regime) are zero-modes of ${\cal D}^s$.

\vspace{2mm}

\nin {\sf Lowering ladder}:  we now wish to compute
\begin{align}
     {\cal D} \Phi_n  = \left({\cal L}_{\bm c}-n \,\sigma_{\bm c}\,  \right)  \Phi_n  .
\end{align}
Proceeding as above we find
\begin{align}
     {\cal D}  \Phi_n &= i{\sf Y}_{\bm l}(\bm \Omega)\,  (\cosh{t})^{-\frac{d-2}{2}} \nonumber \\
     & \times \left(- (z^{2}   -1)\frac{\partial}{\partial z}+\left(n+\frac{d-2}{2}\right)z   \right) P^{-\left( l +\frac{d-2}{2}   \right)}_{n+\frac{d-2}{2}}(z)\, ~~ .
\end{align}
We now use (eqn. 14.10.5 in \cite{dlmf})
$$\left(  -(x^{2}-1)  \frac{  \partial  }{\partial x} + \nu x  \right)P^{-\mu}_{\nu}(x)  =  (\nu - \mu )P^{-\mu}_{\nu-1}(x),$$
and arrive at
\begin{align}
     {\cal D}  \Phi_n  = ~i (n-l)   \Phi_{n-1} .
\end{align}
For the $Q$-modes, the  result is
\begin{align}
     {\cal D}  \tilde\Phi_n  = ~i (n+l+d-2)   \tilde\Phi_{n-1} .
\end{align}
We conclude this section with the following comments:

\vspace{1mm}

\nin i. Physical modes of the $\Delta =-n$ theory ($l\geq n+1$) are mapped onto physical modes of the $\Delta   = - (n-1) $ theory. 

\vspace{1mm}

\nin ii. Gauge modes with $l = n $ of the $\Delta = -n$ theory are zero-modes of $\cal D$. Hence they cannot transform into physical modes under the action of $\cal D$.

\subsection{Vectors}

\subsubsection{Longitudinal modes}

Longitudinal eigenmodes of the vector Laplacian  are well known to be given by derivatives of the scalar harmonics. We do not elaborate on them since they follow easily from the discussion of the previous section \cite{Higuchi:1986},\cite{Kl}.

\subsubsection{Transverse modes}

We now proceed to construct ladder operators for transverse vector harmonics  $A_\mu$ satisfying
\be
\nabla^2 A_\mu= \frac1{\ell^2}\lambda \,A_\mu,~~~\nabla^\mu  A_\mu=0\,,~~~~\lambda= 1 +\Delta\,  \Delta^s 
\label{LV}
\ee
with $\Delta^s$ defined in \eqref{shD} (see App. \ref{LadA} for the allowed eigenvalues on spheres). 

We define two Ladder operators
\begin{align}
     {\cal D}  A_\mu:= \mathcal{L}_{\bm c} A_\mu+(\Delta-1) \sigma_{\bm c} A_\mu-\frac1{\Delta} \nabla_\mu (c^\rho  A_\rho), 
    \label{ladderA}
\end{align}
and
\begin{align}
   {\cal D} ^s A_\mu:=\mathcal{L}_{\bm c} A_\mu+(\Delta^s-1) \sigma_{\bm c} A_\mu-\frac{1}{ \Delta^s} \nabla_\mu (c^\rho  A_\rho) \,.
    \label{ladderAp}
\end{align}
They depend on a CCKV $c^\mu$ and the scale dimension of $A_\mu$, while they also include a compensating \emph{gauge-looking} transformation necessary to preserve the transversality condition. The reader should keep in mind that the gauge-looking transformation becomes an actual invariance only if $\lambda$ is tuned to the strictly massless value (corresponding to $m^{2}=0$ in \eqref{conf-like vectors}). Notice again the similarity of \eqref{ladderA},\eqref{ladderAp} with the infinitesimal form of a conformal vector primary transformation in \eqref{cA}. 

The  action of ${\cal D,D}^s$  on an $\bm A$-mode with scale dimension $\Delta$ gives\footnote{We denote coordinate free 1-forms as $\bm A=A_\mu\,\bm dx^\mu$.}
$$\bm A' :={\cal D}\bm A ~~ \text{and }~ \bm A'' :={\cal D}^s\bm A.  $$
The  modes $\bm A',\bm A''$ have  scale dimensions shifted by one unit
$$\Delta'=\Delta+1 ~~ \text{and } ~\Delta''=\Delta-1.$$ 
These expressions generalize \eqref{shiftSc} to the case of vectors.

---

\nin {\footnotesize{\sf Details and proofs}: consider the ansatz 
\begin{align}
    {\cal D} A_\sigma:=\mathcal{L}_{\bm c} A_\sigma+\gamma\, \sigma_{\bm c} A_\sigma+\alpha\, c^\mu \nabla_\sigma A_\mu 
    \label{ladder}
\end{align}
The coefficients $\gamma$ and $\alpha$ can be fixed by demanding the transformed function to satisfy 
\be
\text{\sf Eigenmode}:~~~\nabla^2({\cal D }A_\sigma)=\frac 1{\ell^2}(\lambda+\delta\lambda)\,{\cal D} A_\sigma
\label{eM}
\ee
with $\delta\lambda$ the change in the eigenvalue, and  
\be
    \text{\sf Transversality}: ~~~~
  0 =\nabla^\sigma ({\cal D} A_\sigma).
\label{trans}
\ee
The need for a compensating ``gauge-looking transformation'' can be seen by taking the divergence of \eqref{ladder}. Indeed,  
\begin{align}
0&=\nabla^\sigma\big(c^\mu\nabla_\mu A_\sigma+ (1+\gamma)\sigma_{\bm c} A_\sigma+\alpha\,c^\mu\nabla_\sigma A_\mu\big)\nn\\
&=c^\mu\nabla^\sigma\nabla_\mu A_\sigma-\frac{1}{\ell^2}(1+\gamma-\alpha\lambda)c^\mu A_\mu
\nn\\
&=\frac{d-1}{\ell^2}c^\mu  A_\mu-\frac{1}{\ell^2}(1+\gamma-\alpha\lambda)c^\mu A_\mu,~~~\leadsto~~~~ \gamma=d-2+\alpha(1+\Delta \Delta^s).
\label{gm2}
\end{align}
To go from the first to the second line we use \eqref{CCKV}, \eqref{LV}, \eqref{cckv3} and \eqref{cckv4}. To get the third line we used that in MSS $[\nabla_\sigma,\nabla_\mu ]A^\sigma= \frac{d-1}{\ell^2}A_\mu$.  The eigenmode condition \eqref{eM} implies
\begin{equation}
    [\ell^2\nabla^2, {\cal  D}]A_\sigma= \delta \lambda\, {\cal D} A_\sigma.
    \label{comu}
\end{equation}
Alternatively computing the lhs  for a MSS space with $\bm c$ a  CCKV one obtains
\begin{align}
[\ell^2\nabla^2, {\cal D}] A_\sigma&=  \underbrace{\left(d-2-2 (\gamma+1)+2 \alpha\right)}_{\delta\lambda\ }  c^\mu\nabla_\mu A_\sigma \nn\\
& +\underbrace{\left(2 \lambda -d(\gamma+1) +2  (d-1) \alpha\right)}_{\delta\lambda\,(1+\gamma)\, } \sigma_{\bm c} A_\sigma\nn \\
& + \underbrace{\left(2+(d-2)\alpha \right)}_{\delta\lambda\, \alpha  } c^\mu\nabla_\sigma A_\mu.
\end{align}
Below the underbraces, we have written what each term should equal for \eqref{comu} to be satisfied. Since $\bm c$ is a CCKV,  using \ref{CCKV}, \ref{cckv2}, and \ref{cckv3} we  get an overdetermined linear system of equations, i.e. three equations for two unknowns $\delta\lambda,\alpha$ ($\gamma$ is given by \eqref{gm2})
\begin{align}
     & {d-2} - {2}  (\gamma+1)+ {2}  \alpha=\delta\lambda\nn \\
     &2\lambda -d(\gamma+1) +2(d-1) \alpha = (1+\gamma) \ \delta\lambda\nn\\
     &2+ (d-2) \alpha =\alpha \ \delta\lambda.
    \label{tosolve}
\end{align}
Amusingly, this system of equations has two solutions corresponding to the rising and lowering ladder operators! The two solutions are
\begin{align}
     \left(\gamma,\alpha,\delta\lambda \right)=\left\{\begin{array}{l}
    \text{I}.\, \left(\frac{\Delta^2-\Delta-1}{\Delta},-\frac{1}{\Delta},d-2-2\Delta \right) \\
    \text{II}.~  \left(\frac{(\Delta^s)^2-\Delta^s-1}{\Delta^s},-\frac{1}{\Delta^s},d-2-2\Delta^s \right).
     \end{array}\right.
     \label{options}
\end{align}
Inserting these solutions in \eqref{ladder} we obtain \eqref{ladderA}-\eqref{ladderAp}.

}

---

\nin {\sf Comment on conformal symmetry}: as for scalar fields, we have  $\delta\lambda=0$ when the scale dimension coincides with the classical value, i.e. $\Delta=\frac d 2-1$. This means that the action of the ladder operators \eqref{ladderA} preserves the space of solutions \eqref{LV}. It is satisfying to recognize that in $d=4$ where the classical engineering dimension is $\Delta=1$ we have
$$\nabla^2A_\mu=\frac3{\ell^2}A_\mu,~~~\nabla_\mu A^\mu=0.$$
These are Maxwell's equations in Lorenz gauge (see \cite{DN},\cite{Higuchi:1987hw},\cite{DW1}).  Moreover, for $\Delta=1$ the ladder $\cal D$ reduces to the Lie action plus a compensating gauge transformation. Hence, the fact that the action of the ladder operators preserves the solution space simply expresses the conformal invariance of Maxwell's equations in $d=4$\footnote{Confront the discussion around eqn.(15) in \cite{Jkw80}.}. 

On the other hand, for  $d \ne 4$, vector fields   with $\Delta = \frac{d}{2}-1$ are massive but still enjoy a  symmetry generated by CCKV. This means that the space of solutions of 
\begin{align}\label{conf-like vectors}
\nabla^2 A_\mu= \frac1{\ell^2}\Bigg( d-1+\underbrace{\frac{(d-2)(d-4)}{4}}_{m^2} \Bigg) \,A_\mu,~~~\nabla^\mu  A_\mu=0
\end{align}
is preserved under the action of 
\begin{align*}
    \mathcal{D}|_{\Delta =\frac{d}{2}-1}   A_{\mu}=   c^{\rho} \left( \nabla_{\rho}  A_{\mu}   - \frac{2}{d-2}  \nabla_{\mu}A_{\rho} \right)  +   \frac{d(d-4)}{2(d-2)}\sigma_{\bm c}\, A_{\mu}.
\end{align*}
Moreover, for $d \geq 4$, the spin-1 fields in (\ref{conf-like vectors}) satisfy the Higuchi bound $m^{2} \geq 0$ (\ref{HB}) and, thus, they are unitary. See the discussions below (around (\ref{HB})) for more details on the Higuchi bound.

For completeness, we quote the commutation relations with the Laplace+ Beltrami operator for arbitrary $d$ and $\Delta$
\begin{align}
    [{\ell^2}\nabla^2, {\cal D}]A_\mu  &=  (  d-2-2\Delta)\,{\cal D }A_{\mu }, \nn\\
    [{\ell^2}\nabla^2, {\cal{D}}^s]A_{\mu } &=  (d-2-2\Delta^s)\,{\cal{D}}^s A_\mu \,.
    \label{DAA}
\end{align}

\nin {\sf Comment on  Killing vectors and ladder operators}: Killing vectors $k_\mu$ are annihilated by $\cal D$.  Comparing  \eqref{nkil} with \eqref{LV}, we learn that Killing vectors have $\Delta=-1$. Inserting this value in \eqref{ladder} and  using \eqref{options},  we obtain
\begin{align}
    {\cal D} k_\mu&= {\cal L}_{\bm c}k_\mu  - \sigma_{\bm c}  k_\mu + c^\rho\nabla_\mu  k_\rho \nn\\
    &=c^\rho (\cancel{\nabla_\rho k_\mu+\nabla_\mu k_\rho})+\nabla_\mu c^\rho A_\rho - \sigma_{\bm c}  k_\mu  =0 .
    \label{KilAn}
\end{align}
The last two terms cancel by virtue of  \eqref{CCKV}. The  result \eqref{KilAn} is the generalization of $  {\cal D}\lvert_{_{\Delta=0}}\Phi_0=0$, shown in figures \ref{SpH} and \ref{Delta diagram}, to the case of vector fields. In passing we comment that there is a tower of shift-symmetric tachyonic vector fields in de Sitter \cite{hinter}. The lowest level field has Killing vectors as zero-modes. However, they give non-unitary representations.

\subsection{Gravitons}

To conclude we consider the case of a symmetric transverse traceless 2-tensors (STT) 
$$ \text{\sf Spin-2}:~~~h_{\mu \nu}=h_{\nu \mu},~~~~ \nabla_\mu h^{\mu \nu}=0 ,~~~ h^\mu_{\ \mu}=0 $$
satisfying
\begin{align}
    \nabla^2 h_{\mu \nu} &=\frac{1}{\ell^2}\lambda \,h_{\mu\nu},~~~~~\lambda=  2+ \Delta  \,\Delta^s  .
    \label{grav}
\end{align}
Ladder operators ${\cal D,D}^s$ depend on a CCKV $c^\mu$ of the positive curvature  MSS, include a compensating \emph{gauge-looking} transformation and take the form\footnote{We denote symmetrization by $B_{(\mu \nu)}:=\frac12 (B_{\mu \nu}+B_{\nu \mu})$.}
\begin{align} 
\label{Mati's}
   {\cal{D}} h_{\mu \nu} &:= \mathcal{L}_{\bm c} h_{\mu \nu}+ (\Delta-2) \sigma_{\bm c} h_{\mu \nu}-  \frac{2}{ \Delta+1}  \nabla_{(\mu}  ( h_{\nu) \rho}\, c^\rho ), \nn \\
   {\cal{D}}^s h_{\mu \nu} &:= \mathcal{L}_{\bm c} h_{\mu \nu}+ (\Delta^s-2) \sigma_{\bm c} h_{\mu \nu}-  \frac{2}{ \Delta^s+1}     \nabla_{(\mu} (  h_{\nu) \rho}\, c^\rho).
\end{align}
The commutation relations with the Laplace+Beltrami operator are
\begin{align}
    &[{\ell^2}\nabla^2, {\cal{D}}]h_{\mu \nu} =  (d-2-2\Delta )h_{\mu \nu} \nn\\
    &[{\ell^2}\nabla^2, {\cal{D}}^s]h_{\mu \nu} =  (d-2-2\Delta^s ) h_{\mu \nu},
    \label{Dh}
\end{align}
which imply that $\cal D$ shifts as $\Delta \leadsto \Delta+1$, while ${\cal D}^s$ lowers the scale dimension by one unit, $\Delta \leadsto \Delta-1$.

---

\nin {\footnotesize{\sf Details and proofs}: we start with
\begin{equation}
     {\cal{D}} h_{\mu \nu}=c^\rho \nabla_\rho h_{\mu \nu}+ (\gamma+2) \,\sigma_{\bm c} h_{\mu \nu}+ \alpha\, c^\rho \nabla_{(\mu} h_{\nu) \rho}.
    \label{eq}
\end{equation}
Again   by taking the divergence of \eqref{eq}, one finds
\begin{equation}
 \nabla^\mu h_{\mu \nu}=0 \leadsto \gamma = \frac{2 \alpha -2 d+4}{(\Delta +1) (d-\Delta )}   .
\end{equation}
To arrive at \eqref{Dh}, we start by computing
\begin{align}
[\nabla^2,   \nabla_\rho] h_{\mu \nu} &=  \nabla^\sigma \nabla_\sigma \nabla_\rho h_{\mu \nu}-\frac\lambda{\ell^2}   \nabla_\rho  h_{\mu \nu}\nn\\& = -\frac\lambda{\ell^2}   \nabla_\rho  h_{\mu \nu}+ \nabla^\sigma \left(\nabla_\rho \nabla_\sigma h_{\mu \nu}+R_{\mu \alpha \sigma \rho} h^\alpha_{\ \nu}+R_{\nu \alpha \sigma \rho} h_\mu^{\ \alpha} \right)\nn\\
&=-\frac{\lambda}{\ell^2} \nabla_\rho h_{\mu \nu}+  \nabla^\sigma \nabla_\rho \nabla_\sigma h_{\mu \nu}+\frac1{\ell^2}(  \nabla_\mu   h _{\rho \nu} +  \nabla_\nu   h _{\rho \mu} )\nn\\
&= \frac1{\ell^2}(  \nabla_\mu   h _{\rho \nu} +  \nabla_\nu   h _{\rho \mu} )  + R_{\sigma\alpha}{}^\sigma{}_\rho\nabla^\alpha h_{\mu\nu}+R_{\mu\alpha}{}^\sigma{}_\rho\nabla_\sigma h^\alpha{}_{ \nu}+R_{\nu\alpha}{}^\sigma{}_\rho\nabla_\sigma h_{ \mu}{}^\alpha \nn\\
&=\frac{d-1}{\ell^2} \nabla_\rho h_{\mu \nu} +\frac4{\ell^2}  \nabla_{(\mu}   h _{\nu)\rho} ,
\end{align}
where we used the first identity in \eqref{MSS} and the transversality condition. Using this expression we obtain
\begin{align}
[\nabla^2, c^\rho \nabla_\rho] h_{\mu \nu} &= (\nabla^2 c^\rho)  \nabla_\rho h_{\mu \nu}+ 2 (\nabla_\sigma c^\rho) (\nabla^\sigma \nabla_\rho h_{\mu \nu})+ c^\rho [\nabla^2, \nabla_\rho] h_{\mu \nu} \nn\\
&=\frac{d-2}{\ell^2}c^\rho\nabla_\rho h_{\mu \nu}+\frac{2\lambda}{\ell^2} \sigma_{\bm c}   h_{\mu \nu}+\frac4{\ell^2} c^\rho \nabla_{(\mu}   h _{\nu)\rho} .
\end{align}
Now, the commutator of the Laplacian with the second term in \eqref{eq} gives
\begin{align}
    [\nabla^2, \sigma_{\bm c}] h_{\mu \nu} &= (\nabla^2 \sigma_{\bm c})\, h_{\mu \nu}+2 (\nabla_\sigma \sigma_{\bm c} )(\nabla^\sigma h_{\mu \nu})\nn\\
    &=- \frac{d}{\ell^2} \sigma_{\bm c} \, h_{\mu \nu}-\frac{2}{\ell^2} c^\sigma \nabla_\sigma h_{\mu \nu}\,.
\end{align}
Finally, to find the commutator with  the last term in \eqref{eq} we compute
\begin{align}
    [\nabla^2, c^\rho \nabla_\mu ]h_{\rho \nu} &=(\nabla^2 c^\rho) \nabla_\mu h_{\rho \nu}+2 (\nabla^\sigma c^\rho)( \nabla_\sigma \nabla_\mu h_{\rho \nu})+c^\rho [\nabla^2, \nabla_\mu] h_{\rho\nu}\nn \\
    &=\frac{d-2}{\ell^2} c^\rho \nabla_\mu h_{\rho \nu}+2 \sigma_{\bm c} \nabla^\rho \nabla_\mu h_{\rho \nu}+\frac{2}{\ell^2} c^\rho (\nabla_{\rho}   h _{\nu\mu}+\nabla_{\nu}   h _{\rho\mu})\nn\\
    &=\frac{d}{\ell^2} c^\rho \nabla_\mu h_{\rho \nu}+2 \sigma_{\bm c} (R_{\rho\alpha}{}^\rho{}_\mu h^\alpha{}_{  \nu}+R_{\nu\alpha}{}^\rho{}_\mu h_\rho{}^\alpha )+\frac{2}{\ell^2} c^\rho \nabla_{\rho}   h _{\nu\mu}-\frac{2}{\ell^2} c^\rho \nabla_{[  \mu}   h _{\nu]\rho}\nn\\
    &=\frac{2}{\ell^2} c^\rho \nabla_{\rho}   h _{\nu\mu}+\frac{d}{\ell^2} c^\rho \nabla_\mu h_{\rho \nu}+2\frac{d}{\ell^2} \sigma_{\bm c}   h _{\mu  \nu} -\frac{2}{\ell^2} c^\rho \nabla_{[  \mu}   h _{\nu]\rho},
\end{align} 
where we used the transversality and traceless condition in going to the third and fourth lines. The  $(\mu\nu)$-symmetrization  results in
\begin{align}
[\nabla^2, c^\rho \nabla_{(\mu}] h_{\nu) \rho}&=\frac{2}{\ell^2} c^\rho \nabla_{\rho}   h _{\mu\nu}+\frac{d}{\ell^2} c^\rho \nabla_{(\mu} h_{\nu)\rho }+\frac{2d}{\ell^2} \sigma_{\bm c}   h _{\mu  \nu}\,. 
\end{align}
Putting everything together and collecting one finds 
\begin{align}
     [{\ell^2}\nabla^2, {\cal D}]h_{\mu \nu}&= \underbrace{\Bigg( d-2-2(\gamma+2)+2 \alpha \Bigg)}_{\delta\lambda} c^\rho \nabla_\rho h_{\mu \nu}\nn \\
     &+ \underbrace{\Bigg(2 \lambda -d (\gamma+2)+2 \alpha d \Bigg)}_{(\gamma+2) \delta\lambda} \sigma_{\bm c} h_{\mu \nu}+ \underbrace{\Big(4+\alpha d  \Big)}_{\alpha \delta\lambda} c^\rho \nabla_{(\mu} h_{\nu) \rho} \,,
\end{align}
where, in the rhs, we have the desired results below the underbraces. This gives an overconstrained system of three equations  for two unknowns $\delta\lambda,\alpha$. As before it has non-trivial solutions, they  are
\begin{align}
(\gamma,\alpha,\delta\lambda)=\left\{
    \begin{array}{l}
    \text{I}.~\left(\frac{\Delta^2-\Delta-4} {\Delta+1},-\frac{2}{\Delta+1},d-2-2\Delta \right) \\
     \text{II}.~ \left(\frac{(\Delta^s)^2-\Delta^s-4} {\Delta^s+1},-\frac{2}{ \Delta^s+1},d-2-2\Delta^s  \right).
     \end{array}\right.
\end{align}
Substituting these values of $\gamma, \alpha, \delta  \lambda $ into \eqref{eq}, and massaging, we find \eqref{Mati's}.
}

---

\nin {\sf Comment on Killing+St\"ackel tensors and ladder operators}: let us show  that Killing+St\"ackel tensors are annihilated by the $\cal D$-operator.  

We start from the definition of a (symmetric) Killing Tensor \cite{pen,pen2,marco,lind}
\begin{equation}
  \text{\sf Killing Tensor}:~~~~  \nabla_\mu k_{\nu \rho}+\nabla_\nu k_{\rho\mu}+\nabla_\rho k_{\mu \nu}=0,
\end{equation}
which is assumed to be traceless.  Contracting with $\nabla^{\mu }$, commuting covariant derivatives and using \eqref{MSS}, one finds
\begin{equation}
    \nabla^2 k_{\mu \nu}=-\frac{2  d  }{\ell^2}   k_{\mu \nu}.
\end{equation}
Hence, STT Killing tensors are solutions of the Laplace+Beltrami operator with $\Delta=-2$. The action of $\cal D$ is given by \eqref{Mati's}, and thus 
\begin{align}
    {\cal D} k_{\mu \nu}&= \mathcal{L}_{\bm c} k_{\mu \nu} -4 \sigma_{\bm c}\, k_{\mu \nu}+2 \nabla_{(\mu}  ( k_{\nu) \rho}\, c^\rho ) \nn \\
    &=c^\rho (\nabla_\rho k_{\mu \nu}+\nabla_\mu k_{\nu \rho}+\nabla_\nu k_{\mu \rho})=0. \nn
\end{align}
There exists also a tower of tachyonic shift-symmetric spin-2 fields in de Sitter, the lowest level of which have STT Killing tensors as zero-modes. They are also   non-unitary (see \cite{hinter} for further details). 

\nin {\sf Comments on conformal-like symmetries and partially massless gravitons}: as for scalars and vectors, the ladder operator $\cal D$ preserves the space of solutions of the spin-2 field equation \eqref{grav} if  $\Delta=\frac{d}{2}-1$. We call these STT {\it conformal-like  gravitons}, they satisfy 
\be
\nabla^2 h_{\mu \nu} =\frac1{\ell^2}\Big( 2+\underbrace{\frac{d}{4}(d-2) }_{m^2}\Big)h_{\mu \nu}.
\label{PMG}
\ee
We should confront this equation with the conditions for having a unitary spin-2 field  in de Sitter space. As found in \cite{Higu2, Higuchi:1986}, STT fields of spin-$s$ satisfying
$$\left(-\nabla^2+s+(2-s)(s+d-3)+m^2\right)h_{\mu_1...\mu_s}=0$$
$$\nabla^\mu h_{\mu\mu_2...\mu_s}=0,~~~h^\mu{}_{\mu\mu_3...\mu_s}=0,$$
give rise to (massive) unitary theories for 
\be
\ell^2m^2>(s -1)(s + d - 4).
\label{HB}
\ee
However, for the special values \cite{DW1,Higu2,DW2,DW3,DW4}
\be
\ell^2m^2 = (\tau- 1)(2s + d- 4 - \tau), ~~~\tau = 1, ..., s ~,
\label{PM}
\ee
the theory is also unitary, by virtue of enjoying a certain gauge invariance. Spin-$s$ fields with masses taking the discrete values given by \eqref{PM} are known as partially massless field of
depth $\tau$\footnote{The case $\tau=1$ corresponds to the  strictly massless case.  In 4d, a strictly massless field has two propagating degrees of freedom with helicities $\pm s$, while a partially massless field of depth $\tau$ has $2\tau$ of them: $(\pm s, \pm(s-1), ..., \pm(s-\tau+ 1))$. Strictly massless fields are the closest analogs of Minkowskian massless fields, while partially massless fields of depth $\tau  > 1$
have no Minkowsian counterparts.}. For the case of spin-2 particles, partially massless gravitons arise for $\ell^2m^2=d-2$\footnote{The gauge invariance of partially massless gravitons which eliminates the zero helicity component of the field is $\delta h_{\mu \nu}=(\nabla_{(\mu}\nabla_{\nu)}+\frac{1}{\ell^{2}}g_{\mu \nu})a$, where $a(x)$ is a scalar gauge parameter.}. We then conclude that in $d\ge4$ the condition \eqref{HB} is satisfied and hence the field equation gives rise to a unitary theory. To conclude, we remark that demanding the gravitons to be conformal-like (in the sense of \eqref{PMG}) and partially massless at the same time is only possible for $d=4$ \cite{Letsios:2023tuc}. Indeed, for the partially massless gravitons ($\Delta =\tfrac d 2 -1 =1$) in $d=4$, the operator (\ref{Mati's}) takes the form
\begin{align}\label{unconv conf sym}
{\cal{D}} h_{\mu \nu}  = c^{\rho}\left(\nabla_{\rho}h_{\mu \nu} - \nabla_{(\mu}h_{\nu)   \rho}  \right),
\end{align}
which coincides with the unconventional conformal symmetry of partially massless graviton uncovered earlier in \cite{Letsios:2023tuc}.

\section{Conclusions}

In the present paper, we have built ladder operators relating distinct discretely labelled UIRs on S$^d$ and dS$_d$. These ladder operators were constructed relying on CCKV on those spaces and they were explicitly written for spins $s=0,1,2$ with generalizations to higher integer spins being straightforward. Our results generalized previous known formul\ae{}  for scalar fields \cite{Cardo,Muck}.

The operators ${\cal D,D}^s$ we found connect solutions to free field equations in S$^d$ and dS$_d$. It is well known that the mass in these equations is related to the quadratic Casimir of the isometry group of the space, therefore, our ladder operators are mass-shifting operators. It is interesting to notice that the form of the ladder operators resembles the spinning conformal primaries transformation laws.

Our setup hopefully sheds light on the appearance of conformal-like symmetries for partially massless gravitons in 4-dimensional de Sitter space \cite{DW5,Barnich:2015tma,Letsios:2023tuc}. Moreover, for $d \neq 4$, we found that massive spin-1 and spin-2 fields with special tunings of their mass parameters ($\Delta = d/2 -1$) enjoy conformal-like symmetries  in the sense that the operator $\mathcal{D}$ which preserves their solution space is built from CKV. The case with $d=4$ is special in the sense that spin-1 and spin-2 fields with $\Delta=1$ are gauge potentials corresponding to the Discrete Series UIRs of SO$(4,1)$, i.e. the Maxwell field and the partially massless graviton on dS$_{4}$. The former is well-known to be SO$(4,2)$ invariant, however, the algebra closure of  the partially massless graviton conformal-like symmetry  is currently under scrutiny \cite{Letsios:2023tuc}.

We leave for future work the computation of the algebra generated by our ladder operators and the possibility of realizing the conformal-like symmetries at the Lagrangian level.

\section*{Acknowledgements}

We would like to thank the participants of the {\it Holography in \& beyond the AdS paradigm} Workshop for creating a stimulating environment that helped start this work. We would like to thank A. Higuchi and F.F. John  for helpfull discussions. This work was funded by CONICET grant  PIP-UE$ 084$,  UNLP grant $X791$ and  PICT 2020-03826. MS is supported by a CONICET fellowship. The work of V. A. Letsios was supported by the Eleni Gagon Survivor's Trust.
  
\appendix
\section{Positive curvature MSS and CCKV}
\label{deSitApp}

\nin {\sf de Sitter}:  in global conformal coordinates the dS$_4$ metric is given by
\begin{equation}
    ds^2= \frac{1}{\sin{t}^2}(-dt^2+ d\chi^2+\sin^2{\chi}( d\theta^2+  \sin{\theta}^2 d\phi^2)
    \label{metric}
\end{equation}
here $t\in(-\pi,0)$. Notice this parametrization maps dS$_4$ to a portion of Einstein Static Universe \cite{HE}. 

The Killing Vectors on dS$_4$ can be separated as those acting closely on $t=const$ (spatial sections),   $\{\bm J_i,\bm P_i\}$, which close a SO$(4)$ algebra, and $\{\bm D,\bm K_i\}$ which move us away from the $t=const.$ hypersurface. Their explicit form is
\begin{align}
\bm J_1 &=  \sin\phi \,\bm\partial_\theta + \cot \theta\cos \phi  \,\bm\partial_\phi\nn\\  
\bm J_2 &=-\cos\phi \,\bm \partial_\theta + \cot \theta\sin \phi  \,\bm \partial_\phi\nn\\    
\bm J_3 &= -\bm\partial_\phi
\label{Js}
\end{align} 
\begin{align}
&\bm P_1=   \sin\theta \cos\phi \,\bm \partial_\chi+  \cot \chi  \cos \theta \cos \phi\, \bm \partial_\theta -\cot \chi  \csc\theta \sin\phi \,\bm \partial_\phi\nn\\
&\bm P_2=   \sin\theta \sin\phi \,\bm \partial_\chi+  \cot \chi  \cos \theta \sin \phi\, \bm \partial_\theta +\cot \chi  \csc\theta \cos\phi \,\bm \partial_\phi\nn\\
&\bm P_3=\cos\theta \,\bm \partial_\chi- \cot \chi\sin\theta \, \bm \partial_\theta \, , 
\label{Ps}
\end{align} 
\begin{align} 
\bm D = &\sin t \cos\chi \,\bm \partial_t +\cos t \sin \chi\, \bm \partial_\chi\nn\\
\bm K_1= & -\sin t\sin\chi \sin\theta \cos\phi \,\bm\partial_t +\cos t\cos\chi\sin\theta\cos\phi \,\bm \partial_\chi\nn\\
&+ \cos t \csc \chi  \cos \theta \cos \phi\, \bm \partial_\theta -\cos t\csc \chi  \csc\theta \sin\phi \,\bm \partial_\phi\nn\\
\bm K_2= &  -\sin t\sin\chi \sin\theta \sin\phi \,\bm\partial_t +\cos t\cos\chi\sin\theta\sin\phi \,\bm \partial_\chi\nn\\
&+ \cos t \csc \chi  \cos \theta \sin \phi\, \bm \partial_\theta +\cos t\csc \chi  \csc\theta \cos\phi \,\bm \partial_\phi\nn\\
{\bm K}_3=&-\sin t\sin\chi\cos\theta\,\bm\partial_t+\cos t\cos\chi\cos\theta \,\bm \partial_\chi- \cos t\csc \chi\sin\theta \, \bm \partial_\theta \, .  
\label{boost}
\end{align}
The set of Killing vectors $\{\bm J_i,\bm P_i,\bm D,\bm K_i\}$ with $i=1,2,3$ close a SO(4,1) algebra  
$$[\bm J_i,\bm J_j]=\epsilon_{ijk}\,\bm J_k,~~~~[\bm J_i,\bm P_j]=\epsilon_{ijk}\bm P_k,~~~~[\bm J_i,\bm K_j]=\epsilon_{ijk}\bm K_k$$
$$[\bm P_i,\bm P_j]=\epsilon_{ijk}\,\bm J_k,~~~~[\bm K_i,\bm K_j]= \epsilon_{ijk}\bm J_k$$
\be
[\bm D,\bm P_i]=-\bm K_i,~~~~[\bm D,\bm K_i]=-\bm P_i,~~~~[\bm P_i ,\bm K_j]=-\bm D\,\delta_{ij}\,.
\label{so41}
\ee
In addition dS$_4$ posses five proper CKV which we write
\begin{align}{\bm{\Tilde{D}}}=&\nonumber \cos t \cos \chi  \bm\partial_t-\sin t \sin \chi  \bm\partial_\chi
    \\ 
 {\bm{\Tilde{K}}}_1=&  \sin \theta  \cos t \sin \chi  \cos \phi\, \bm\partial_t+\sin \theta  \sin t \cos \chi  \cos \phi\, \bm\partial_\chi \nn\\ 
&+\cos \theta  \sin t \csc \chi  \cos \phi\,  \bm\partial_\theta-\csc \theta  \sin t \csc \chi  \sin \phi\, \bm\partial_\phi \nonumber \\{\bm{\Tilde{K}}}_2=&\nonumber\sin  \theta   \cos t \sin \chi  \sin \phi  \, \bm\partial_t+\sin \theta \sin t  \cos \chi  \sin \phi  \,\bm\partial_\chi \\ 
&+\cos \theta  \sin t  \csc  \chi   \sin  \phi   \, \bm\partial_\theta+\csc \theta  \sin t  \csc \chi  \cos \phi \ \bm\partial_\phi \nonumber\\
{\bm{\Tilde{K}}}_3=&\cos \theta  \cos t \sin \chi \bm\partial_t+\cos \theta  \sin t \cos \chi  \bm\partial_\chi-\sin \theta  \sin t \csc \chi  \bm\partial_\theta\nn\\
{\bm{\Tilde{T}}}=&-\bm\partial_t 
\label{CKV}
\end{align}
The conformal symmetry algebra of dS$_4$ is SO(4,2) which in addition to \eqref{so41} reads
$$[\Tilde{\bm K}_i,\bm \Tilde{\bm K}_j]= -\epsilon_{ijk}\bm J_k,~~~~[\bm\Tilde{\bm K}_i,\bm K_j]=\Tilde{\bm T},~~~~ [\bm P_i, \bm \Tilde{\bm K}_j]=\bm \Tilde{\bm D}\, \delta_{ij},~~~~[\bm D,\bm\Tilde{\bm K}_i]=0$$ $$[\bm D,\Tilde{\bm D}]=\Tilde{\bm T},~~~~[\bm \Tilde{\bm D},\bm\Tilde{\bm K}_i]=\bm P_i,~~~~[ \bm \Tilde{\bm D},\bm P_i]=\tilde{\bm K}_i,~~~~[\bm J_i, \tilde{\bm K_j}]=\epsilon_{ijk} \tilde{\bm{K}_k}$$ $$[\bm D,\tilde{\bm T }]=\tilde{\bm D},~~~~[\tilde{\bm K}_i,\tilde{\bm T}]=\bm K_i,~~~~[\bm K_i, \tilde{\bm T}]=-\tilde{\bm K}_i$$ $$[\Tilde{\bm D},\bm J_i]=[\Tilde{\bm D},\bm P_i]=[\Tilde{\bm T},\bm J_i]=[\Tilde{\bm T},\bm P_i]=0$$
The proper CKV \eqref{CKV} were chosen to be closed,   if we pull down the index one can verify that all each of them can be written $c_\mu=\partial_\mu\Psi_c$ (cf.\eqref{clo}). The  potentials  are
\begin{align}
\bm{ \Tilde{D} } :& ~~~\Psi_D= \csc t \cos \chi  \nonumber \\
\bm{\tilde  K_1  }:& ~~~\Psi_{K_1}=\sin \theta  \csc t \sin \chi  \cos \phi   \nonumber \\ \bm{\Tilde{K_2}} :&~~~\Psi_{K_2}=\sin \theta  \csc t \sin \chi  \sin \phi   \nonumber \\
{\bm{\Tilde{K}}}_3:&~~~\Psi_{K_3}=\cos \theta  \csc t \sin \chi \nonumber \\
{\bm{\Tilde{T}}}:& ~~~\Psi_T=-\cot t  
\label{pD}
\end{align}
It is amusing to verify that the proper CKV potentials satisfy
\be
(\Psi_D)^2+(\Psi_{K_i})^2-(\Psi_T)^2=1
\label{confP}
\ee
This is in agreement with the fact that they coincide with the embedding coordinates $\Psi\sim X^I$ (see \cite{Alleng,Letsios}) when representing de Sitter dS$_d$ as a quadric in Minkowski embedding space $\eta_{IJ}X^IX^J=1$ with $I,J=0,...,d$ .  

\vspace{2mm}

\nin{\sf Sphere}: the $S^3$ metric   is written as
\begin{equation}
    ds^2=  d\chi^2+\sin^2{\chi}( d\theta^2+  \sin{\theta}^2 d\phi^2)
    \label{metric2}
\end{equation}
The isometries   are generated by $\{\bm J_i,\bm P_i\}$ closing a SO(4) algebra
$$[\bm J_i,\bm J_j]=\epsilon_{ijk}\,\bm J_k,~~~~[\bm J_i,\bm P_j]=\epsilon_{ijk}\bm P_k,~~~~ [\bm P_i,\bm P_j]=\epsilon_{ijk}\,\bm J_k\,.$$
The proper CKV on the sphere can be obtained from the late-time limit behavior of dS$_4$ Killing vector \emph{boosts} \eqref{boost}. Combined with  Their covariant expressions are given by
\begin{align}
    \bm D &= \bm d(-\cos \chi)\nn \\
    \bm K_1 &= \bm d(\sin \chi \sin \theta \cos \phi) \nn\\
    \bm K_2 &= \bm d (\sin \chi \sin \theta \sin \phi) \nn\\
    \bm K_3 &= \bm d (\sin \chi \cos \theta)
\label{pS}
\end{align}
with $\bm d=\bm dx^\mu\,\partial_\mu$ the exterior differential operator. In analogy with \eqref{confP}, on the sphere,  one can check that the proper CKV potentials $\Psi_I$ ($I=\{D,K_i\}$) satisfy 
$$(\Psi_D)^2+(\Psi_{K_i})^2 =1
$$

\section{Scalar harmonics on $S^{d-1}$}
\label{SphArm}

We write the metric of $S^{d-1}$  in standard fashion
\be 
d\Omega_{d-1}^2=d\theta_{d-1}^2+\sin^2\theta_{d-1} \,d\Omega_{d-2}^2\,.
\label{Smetric}
\ee
The full set of coordinates is $\bm\Omega=(\theta_{d-1},\theta_{d-2},\theta_{d-3},..,\theta_2,\varphi)$ where $\theta_i\in[0,\pi)$ and $\varphi\in[0,2\pi)$.  

The scalar spherical harmonics on $S^{d-1}$ are classified by $(d-1)$ quantum numbers 
\be
\bm l=(l_{d-1},l_{d-2},..,l_1),~~~~  l_i=0,1,2,..
\label{list}
\ee
obeying
$$l_{d-1}   \geq  l_{d-2} \geq ... \geq l_{2}   \geq |l_{1}| \geq 0.$$ 
The first quantum number in the list \eqref{list}, i.e. $l_{d-1}$, parametrizes the Laplace eigenvalue\footnote{We denoted $k=l_{d-1}$ in \eqref{SH} and $l=l_{d-1}$ in app. \ref{SSs}.}  
\be
- {\nabla}_{sph}^{2}   {\sf Y}_{\bm l}  =  l_{d-1} (l_{d-1} + d -2) \, {\sf Y}_{\bm l} ,~~~~~~l_{d-1}=0,1,2,... 
\label{sAr}
\ee
They take the explicit form \cite{Higuchi:1986}
\begin{align} 
 {\sf Y}_{\bm l}(\bm{\Omega}) = \prod_{r = 1}^{d-2}\left( \phantom{\hspace{-3cm}P^{-\left( l_{d-r-1} +\frac{d-r-2}{2}   \right)}_{l_{d-r}+\frac{d-r-2}{2}}}\,C(l_{d-r},l_{d-r-1})\right.\, &\left.\left( \sin{\theta_{d-r}}   \right)^{-(d-r-2)/2}\right.\nn\\
 &\left.P^{-\left( l_{d-r-1} +\frac{d-r-2}{2}   \right)}_{l_{d-r}+\frac{d-r-2}{2}}( \cos{\theta_{d-r}}) \right)  \times   \frac{1}{ \sqrt{2 \pi}} e^{l_{1}  \varphi},
 \label{YP}
\end{align}
where    
\begin{align}  
    C(l_{d-1},l_{d-2}) = \left( \frac{2 l_{d-1}+d-2}{2}  \,   \frac{(l_{d-1}+l_{d-2}+d-3)!}{(l_{d-1}-l_{d-2})!}   \right)^{1   /2 }.
    \label{norma}
\end{align}
In \eqref{YP}, $P_{\nu}^{-\mu}$ are the associated Legendre function of the first kind \cite{Gradshteyn}. The    normalisation factors   ensure that
\begin{align}
    \int_{sph} \left({\sf Y}_{\bm l}(\bm{\Omega})\right)^{*} \,  {\sf Y}_{\bm l'}(\bm{\Omega}) = \delta_{\bm l,\bm l'}.
\end{align}
The harmonics on $S^{d-1}$ can be written in terms of $S^{d-2}$ harmonics 
${\sf Y}_{\hat{\bm l}}(\hat{\bm \Omega})$
where $\hat {\bm l}=(l_{d-2},l_{d-3},..,l_1)$  and $\hat{\bm\Omega}=(\theta_{d-2},..,\theta_2,\varphi)$ 
as
\begin{align}
 {\sf Y}_{\bm l}(\bm{\Omega}) &=    C (l_{d-1},l_{d-2})\, \left( \sin{\theta_{d-1}}   \right)^{-(d-3)/2}\,P^{-\left( l_{d-2} +\frac{d-3}{2}   \right)}_{l_{d-1}+\frac{d-3}{2}}( \cos{\theta_{d-1}}) \, {\sf Y}_{\hat{\bm l}}(\hat{\bm \Omega}).
\end{align}


\section{Shift-symmetric scalars in global dS} 
\label{SSs}

\subsection{Mode solutions for shift-symmetric scalars on dS$_d$}

Shift-symmetric scalars on $dS_{d}$ satisfy (de Sitter radius $\ell^2=1$)
\begin{align}
    \nabla^{2}  \Phi = -  n(n+d-1)  \,   \Phi\,, ~~~~~~~n=0,1,2,...
    \label{sS}
\end{align}
To find the solution to this equation in global coordinates \eqref{glob}  we make the ansatz 
\be
\Phi(t, \bm{\Omega} ) =   f({t}) \, {\sf Y}_{\bm l}(\bm{\Omega}  )~~~~\text{with}~~~{\bm l}=(l,l_{d-2},..,l_1)\,.
\label{AnS}
\ee
Here $ {\sf Y}_{\bm l}(\bm{\Omega})$ are the scalar spherical harmonics on  $S^{d-1}$ (see app. \ref{SphArm}). Inserting the ansatz \eqref{AnS} in \eqref{sS} one finds\footnote{The general solution to \eqref{f} is  
\be
f(t)=\frac1{(\cosh t)^{\frac{d-1}2}}\left(c_1\ P_{ l+\frac{d-3}2}^{ n+\frac{d-1}2}(\tanh t)+c_2\ Q_{ l+\frac{d-3}2}^{ n+\frac{d-1}2}(\tanh t)\right)
\label{f}
\ee
Discrete/Exceptional UIRs follow for particular  choices of $c_{1,2}$ for  $f(t)$. The precise choice of coefficients relating \eqref{f} to \eqref{posE} was originally spelled out in \cite{Tagirov} using the Whipple formula (see sect. 5.5 in \cite{BD} and \cite{HiguSym}).}
\be
f''+(d-1)\tanh t\,f'+\left(-n(n+d-1)+\frac{l(l+d-2)}{\cosh^2t}\right)f=0,
\label{EOMf}
\ee
 where the prime denotes differentiation with respect to $t$.

The appropriate \emph{positive norm modes}  for the quantization of a scalar field in dS were introduced originally in \cite{CT,Tagirov}. For the case of tachyons they are (see also \cite{Higuchi:1986,BD,EM,Ak})
\begin{align}
   \Phi_{\bm l }(t,\bm  \Omega ) =& \, \left( \cosh{t}   \right)^{-(d-2)/2}\,P^{-\left( l  +\frac{d-2}{2}   \right)}_{n+\frac{d-2}{2}}(i \sinh{t})\, ~ {\sf Y}_{\bm l}(\bm{\Omega} ) ,~~~~l> n
   \label{posE}
\end{align}
with the KG inner product, in global coordinates, taking the form
\be
(\Phi,\Psi)=i(\cosh t)^{d-1}\int_{sph}d\bm\Omega \,\left(\Phi^*\,\partial_t\Psi-(\partial_t\Phi)^*\Psi\right).
\label{IP}
\ee
The argument of the Legendre function in \eqref{posE} can be understood as arising from the Wick rotation \eqref{wick} on the $\cos\theta\mapsto i\sinh t$  which appears in the  spherical harmonics \eqref{YP}. 

\vspace{2mm}

\nin Some comments are in order:

\vspace{1mm}

\nin 1.  $P_{\nu}^{-\mu}$ are the associated Legendre function of the first kind which can be expressed in terms of the Gauss hypergeometric function as \cite{Gradshteyn}
\begin{align}
    P^{-\mu}_{\nu}(z) = \frac{1}{\Gamma(\mu  +1)}  \,   \left(   \frac{1-z}{1+z} \right)^{\mu  /2} ~ F\left(    -\nu,  \nu+1 ;  \mu +1 ; \frac{1-z}{2}\right),
\end{align}
thus,
\begin{align}\label{mshyp}
   \Phi_{\bm l}(t, \bm{\Omega} ) =& \, \frac{1}{\Gamma(l  + \frac{d}{2} )} \left( \cosh{t}   \right)^{-(d-2)/2}~  \left(   \frac{1-i \sinh{t}}{1+ i \sinh{t}} \right)^{\frac l2  + \frac{d-2}{4}  }  
   \,\nonumber \\ 
   & \times  F\left(  -n -\frac{d-2}{2}, n+\frac{d}{2} ; l + \frac{d}{2}  ; \frac{1 - i   \sinh{t}}{2} \right)  {\sf Y}_{\bm l} (\bm \Omega).
\end{align}

\vspace{1mm}

\nin 2. In the short wavelength limit, $l  \gg 1$, the modes \eqref{posE} behave as \cite{CT,Higuchi:1991}\footnote{See the discussion around eqn. (5.73) in \cite{BD}.}
\begin{align} 
\label{+E}
   \frac{\partial}{  \partial  t}  \Phi_{\bm l}(t, \bm{\Omega} )  \sim  -il\,\frac{\Phi_{\bm l}(t, \bm{\Omega} )}{ \cosh{t}} \, , ~~~\text{for}~~l   \gg 1,
\end{align}
which is interpreted as a generalized positive frequency behaviour. Notice   $\partial_\tau=\cosh t\,\partial_t$ is the proper time of a comoving observer in de Sitter space.  

\vspace{1mm}

\nin 3. The scalar product, of any two modes  solutions \eqref{posE} is \cite{Higuchi:1986}
\begin{align}
   \left( \Phi_{\bm l},\Phi_{\bm l'}\right) 
     = \frac{2}{\Gamma(l-{n})   \,  \Gamma(l +{n}+d-1)}  \delta_{\bm l,\bm l' } .
\end{align}
This important result leads to the following consequences:

\vspace{2mm}

\nin . \textbf{Physical modes}: correspond to the set of (complex) modes with $l > n$, they have positive norm.\\
. {\bf Gauge modes}: correspond to the set of $ l \leq n$ modes (real up to a phase), they have zero norm.    

\vspace{2mm}

\nin Important features of the  set   $\{\Phi_{\bm l}\}$ built with $P$-functions are: 

\vspace{.5mm}

\nin . Gauge modes   transform among themselves under SO($d,1$) \cite{Higuchi:1986}, we will summarize this below.  As expected, the number of gauge modes coincides with the number of independent symmetric polynomials appearing in \eqref{gm}\footnote{As an example consider $n=1$ in \eqref{sS}. The condition $l\le n$ gives as possible harmonics
$$\bm l \in \{(0,...,0),(1,0,..0),\underbrace{( 1,1,0,..0),(  1,1,1...1)}_{d-2~\text{terms}},(1,1,...,-1)\}$$ 
which comprise $d+1$ distinct gauge modes. This agrees with the $d+1$ embedding space coordinates.}. They are zero-norm modes because for $l\le n$ the associated Legendre $P$-functions appearing in \eqref{posE} are proportional to their complex conjugates (see \eqref{cP} below). \\
. Physical modes transform among themselves and into  gauge modes (see below). It was shown in \cite{Higuchi:1986} that the physical modes form a UIR of SO($d,1$) \cite{Higuchi:1986}.

\vspace{1mm}

\nin 4. For $l>n$ the Wronskians in section 14.2(iv) in \cite{dlmf}   show that the complex conjugates $(\Phi_{\bm  l})^*$ are \emph{negative norm} states
$$ \left( \Phi_{\bm l}^*,\Phi_{\bm l'}^*\right) 
     =- \left( \Phi_{\bm l},\Phi_{\bm l'}\right),~~~~ \left( \Phi_{\bm l}^*,\Phi_{\bm l'}\right) 
     = 0\,.$$

\subsection{Seed modes}

As we mentioned earlier, in the range $l\le n$ the complex conjugates of the $P$-functions in \eqref{posE} are not independent from the $P$-functions (cf. \eqref{cP}). The second set of independent solutions to \eqref{sS} is 
\begin{align} 
   \tilde\Phi_{\bm l}(t, \bm{\Omega} ) =& \, \left( \cosh{t}   \right)^{-(d-2)/2}\,Q^{ l+\frac{d-2}{2}   }_{n+\frac{d-2}{2}}(i \sinh{t})\, {\sf Y}_{\bm l} (\bm \Omega) ,
   \label{Qf}
\end{align}
where $Q^{\mu}_{\nu}(z)$ are the associated Legendre functions of the second kind,  analytic in the cut complex plane \cite{Gradshteyn,dlmf}. There are several alternative representations for the $Q$-functions which will not be relevant to our discussion. For our purposes, we consider $Q^{\mu}_{\nu}(z)$ to be given by the analytic extension, beyond the unit circle, of eqn. 14.3.20 in \cite{dlmf}.  

\nin The $Q$-functions satisfy the following properties:

\vspace{1mm}

\nin i. {\sf Zero norm}: being $\mu,\nu\in\mathbb R$, when evaluated at $z = i \sinh{t}$ we have that   $ \left( Q^{\mu}_{\nu}(z)\right)^{*} =  Q^{\mu}_{\nu}(-z)$. Then, using eqn. 14.24.2 in \cite{dlmf} we conclude that for $Im(z) <0$
$$Q^{\mu}_{\nu}(-z) = -e^{-i \nu \pi}Q^{\mu}_{\nu}(z) ~~~\leadsto~~\boxed{\left( Q^{\mu}_{\nu}(z)\right)^{*}\propto Q^{\mu}_{\nu}(z)}.$$  

\nin ii. {\sf Linear dependence}: for $l>n$, $Q^\mu_\nu$ are customarily disregarded, as they are not independent from $\{P^{-\mu}_{\nu}(i\sinh t),P^{-\mu}_{\nu}(-i\sinh t)\}$. This can be seen from  (eqn 14.24.1 in \cite{dlmf})
\be 
P^{-\mu}_{\nu}(-z)-  e^{- i   \pi   \nu}  P^{-\mu}_{\nu}(z) =     \frac{2  }{\Gamma( \mu - \nu)} \frac{e^{- i \mu \pi }}{\Gamma(\mu +\nu +1 )}Q^{\mu}_{\nu}(z),~~Im(z)>0.
\label{cP}
\ee
A  similar relation holds for $Im(z) <0 $.

\nin iii. {\sf Linear independence}: for $l\le n$, the  set of linear independent solutions to \eqref{sS} is $\{P^{-\mu}_{\nu}(i\sinh t),Q^{\mu}_{\nu}(i\sinh t)\}$. Independence follows from eqn. 14.2.6 in \cite{dlmf}.

\subsection{de Sitter isometry and mode expansions}

In this section, we study the transformation properties of the mode functions $\{\Phi_{\bm l},\tilde\Phi_{\bm l}\}$ in  \eqref{posE} and \eqref{Qf}  under SO$(d,1)$. The transformation properties for the $P$-modes were originally uncovered in \cite{Higuchi:1986}. Here we extend it to the (seed) $Q$-modes \eqref{Qf}. 

--

\nin We start by mentioning that the $l$-quantum number in \eqref{sAr} is preserved under the action of SO$(d)\subset$ SO$(d,1)$. However, under a  SO$(1,1)\subset$ SO$(d,1) $ boost, the $l$-quantum number shifts by one. For concreteness, we write the sphere   metric in \eqref{Smetric} as
$$d\Omega_{d-1}^2=d\vartheta^2+\sin^2\vartheta \,d\Omega_{d-2}^2$$
and consider, without loss of generality, the SO(1,1) de Sitter Killing vector  
\begin{align}  
    \bm\xi  =  \cos{\vartheta }\, \bm\partial_ t   - \tanh{t} \, \sin{\vartheta }\,\bm \partial_  \vartheta .
    \label{KilVec}
\end{align}
Our aim is to study its action on the set of modes 
$$\Psi_{\bm l}(t, \bm{\Omega} ) =   f_l({t}) \, {\sf Y}_{\bm l}(\bm{\Omega}),  $$
where
$$f_l(t)\in \left\{\left( \cosh{t}   \right)^{-(d-2)/2}P^{-\left( l  +\frac{d-2}{2}   \right)}_{n+\frac{d-2}{2}}(i \sinh{t}),\,\left( \cosh{t}   \right)^{-(d-2)/2}Q^{  l  +\frac{d-2}{2}}_{n+\frac{d-2}{2}}(i \sinh{t})\right\}.$$
We show below that \eqref{KilVec} leads to a simple mixing of modes which can be schematically summarized as
$$\bm l=(l,l_{d-2},..,l_1)~~\mapsto~~\bm l'=(l\pm1,l_{d-2},..,l_1).$$
Moreover, the important results are:

\vspace{2mm}
\nin . Gauge modes $l\le n$ ($P$-modes) do not mix with $l>n$ modes (see \eqref{Pmodes}).

\vspace{1mm}

\nin . Seed modes  with $l\le n$ ($Q$-modes) can be connected with $l>n$ (see \eqref{xiQ}). However, from \eqref{cP} we conclude the result involves a combination of positive and negative norm  $P$-modes.

---

{\footnotesize \nin {\sf Details}: to analyze the action of symmetries, following \cite{Higuchi:1986} we re-express the Lie derivative   ${\cal L}_{\bm\xi}$  as
$$  \mathcal{L}_{\bm\xi}   \Psi_{\bm l}  =  \frac{1}{2 l   + d -2} \left(   T^{(+)}   +  T^{(-)}\right) \Psi_{\bm l}  $$
with
\begin{align}
 T^{(+)} = D^{(+)}  \times   \tilde{D}^{(+)},
\end{align}
where
\begin{align}
     D^{(+)} &=   \partial_t -l   \,  \tanh t.\nn\\
     \tilde D ^{(+)} &= \sin \vartheta\, \partial_\vartheta     +(l+d-2) \cos \vartheta,
     \label{Dplus}
\end{align}
and
\begin{align}
    T^{(-)} = D^{(-)}\times\tilde D^{(-)},
\end{align}
where
\begin{align}
     D^{(-)}&=\partial_t +(l+d-2)   \,  \tanh t\nn\\
    \tilde D^{(-)}&= -\sin\vartheta \,  \partial_\vartheta   +l \,   \cos\vartheta.
    \label{Dminus}
\end{align}
The virtue of the decomposition  $T=D\times\tilde D$ is that it factorizes the action on the spatial slices, $\tilde D$, from the action along time, $D$\footnote{It is interesting to notice that the operators $D^{(\pm)},\tilde D^{(\pm)}$ in \eqref{ladercio} are  related to the ladder operators ${\cal D,D}^s$ defined in the main text. For example, ${\cal D} =-\cosh t\, D^{(+)}$, etc. However, while ${\cal D,D}^s$ shift the mass eigenvalue keeping the sphere quantum numbers remain unchanged, the operators $D^{(\pm)},\tilde D^{(\pm)}$ shift the (principal) angular quantum number.}. 

---

\nin Using several properties of the associated Legendre functions one finds: \\
. $\tilde{D}^{(+)}$  raises the principal quantum number $l\mapsto l+1$
\begin{align}
 \tilde{D}^{(+)}{\sf Y}_{\bm l}(\bm{\Omega}) = \, \frac{C(l , l_{d-2})}{C(l+1, l_{d-2})  }  (l+l_{d-2} +d-2)
  \,{\sf Y}_{\bm l^+}(\bm{\Omega}) ,
\end{align}
here $\bm l^+=(l+1,l_{d-2},..,l_1)$ and the normalization factors are defined in \eqref{norma}.

\vspace{1mm}

\nin. $\tilde{D}^{(-)}$  lowers the principal quantum number $l\mapsto l-1$
\begin{align} 
\tilde{D}^{(-)}{\sf Y}_{\bm l}(\bm{\Omega}) = \, \frac{C (l , l_{d-2})}{C(l-1,l_{d-2})  }  (l-l_{d-2} )
  \,{\sf Y}_{\bm l^-}(\bm{\Omega}) ,
\end{align}
here $\bm l^-=(l-1,l_{d-2},..,l_1)$.

\vspace{1mm}

\nin .  $ D^{(+)}$ shifts $l\mapsto l+1$ in the $P$-function
\begin{align}
     D^{(+)}& \left(  (\cosh t)^{-\frac{d-2}{2}} \,P^{-(l  + \frac{d-2}{2} )}_{n + \frac{d-2}{2}}(i\sinh t)  \right) \nonumber\\
    &=  k^{(+)}_{P} \, (\cosh t)^{-\frac{d-2}{2}} \,P^{-(l +1 + \frac{d-2}{2} )}_{n + \frac{d-2}{2}}(i\sinh t) 
\end{align}
where
\begin{align} 
    k^{(+)}_{P} = i(n - l )(n + l  + d -1).
\end{align}
Acting on the Legendre $Q$-functions, one obtains
\begin{align}
     D ^{(+)}& \left(  (\cosh t)^{-\frac{d-2}{2}} \,Q^{l + \frac{d-2}{2}}_{n + \frac{d-2}{2}}(i\sinh t)  \right) \nonumber\\
    &=  k^{(+)}_{Q} \, (\cosh t)^{-\frac{d-2}{2}} \,Q^{l+1 + \frac{d-2}{2} }_{n + \frac{d-2}{2}}(i\sinh t) ,
\end{align}
where 
\begin{align} 
    k^{(+)}_{Q} = 1.
\end{align}
. Proceeding similarly, the $D^{(-)}$ action can be summarized as
\begin{align}
    D^{(-)}    f_l    = k^{(-)}_{f} \,  f_{l-1}  ,
\end{align}
where $k_f^{(\pm)}$ depends on whether the $D^{(-)}$ acts on $P$- or $Q$-functions. The factors $ k^{(-)}_{f}$ are given by
$$k^{(-)}_P=-i ~~~~\text{and}~~~~k^{(-)}_{Q} = (n-l+1)  (n+l+d-2) . $$

\nin Combining the results above we have
\begin{align}
    T^{(+)} \Psi_{\bm l} &=  \frac{C (l , l_{d-2})}{C (l +1,l_{d-2})  }  (l  + l_{d-2} +d-2) \,  k^{(+)}_{f}  \,  \Psi_{\bm l^+} ,  \nn\\
    T^{(-)} \Psi_{\bm l}  &=   \frac{C (l , l_{d-2})}{C (l -1, l_{d-2})  }  (l - l_{d-2} ) \,  k^{(-)}_{f}  \, \Psi_{\bm l^-} .
\end{align}
Long story short,   we conclude that
\begin{align}
    \mathcal{L}_{\bm\xi}   \Psi_{\bm l}& =   \frac{1}{2 l   + d -2} \left(   T^{(+)}   +  T^{(-)}\right) \Psi_{\bm l} \nn \\
    &=  \frac{\,C (l ,l_{d-2}) }{2 l + d -2} \left( \frac{l + l_{d-2} +d-2}{C (l +1,l_{d-2})  }   \,  k^{(+)}_{f}  \, \Psi_{\bm l^+}+   \frac{l   - l_{d-2}}{C (l -1, l_{d-2})  }   \,  k^{(-)}_{f}  \,\Psi_{\bm l^-}\right).
    \label{ladercio}
\end{align}  
}
 
---
  
\nin\textbf{Transformation of $P$-modes}: specialising to $\Phi_{\bm l}$ in \eqref{posE} one obtains, forgetting the factor in front of the parentheses in \eqref{ladercio},
 \begin{align}
     \mathcal{L}_{\bm\xi}   \Phi_{\bm l}\propto&   \frac{l+ l_{d-2} +d-2}{C(l +1, l_{d-2})}\, (n-l)(n+l +d-1)  \,   \Phi_{\bm l^+} - \frac{l  - l_{d-2}}{C(l-1, l_{d-2})  } \, \Phi_{\bm l^-} .
    \label{Pmodes}
\end{align} 
The $(n-l)$ factor multiplying $\Phi_{\bm l^+}$ implies that gauge modes ($l\le n$) cannot be turned into physical modes ($l>n$)  under the action of dS isometries. Concomitantly it follows that all gauge modes ($l\le n$) mix among themselves. However, physical modes can be connected with gauge modes (due to the last term in \eqref{Pmodes}).  

\vspace{2mm}

\noindent   \textbf{Transformation of $Q$-modes}: for the seed modes  \eqref{Qf}  we obtain
\begin{align}
   \mathcal{L}_{\bm\xi}   \tilde\Phi_{\bm l}\propto&  \frac{l + l_{d-2} +d-2}{C(l+1,l_{d-2})  }   \, \tilde\Phi_{\bm l^+}
    + \frac{l-l_{d-2}}{C(l-1, l_{d-2})  }   \, \, (n - l +1)(n +l  + d-2)\,\tilde\Phi_{\bm l^-}.
    \label{xiQ}
\end{align}  
Some comments are in order: contrary to \eqref{Pmodes}, the first term in \eqref{xiQ} never vanishes. This implies that $Q$-modes with $l\le n$ can be connected with physical modes  ($l>n$) under the action of dS isometries. However,  using \eqref{cP} is is important to stress that the result  is a $P$-mode  combination of positive and negative norm.


\section{Vector Spherical harmonics in  $S^3$ and  Ladder operators}
\label{LadA}

Transverse vector harmonics (TVH) on $S^3$ were thoroughly discussed  in \cite{AGS}. We show here how to obtain the full spectrum using ladder operators .

TVH are classified by the eigenvalues of the two Casimirs of SO$(4)$  
\be
\mathcal{C}=\bm P_i^2+\bm J_i^2~~~\text{and}~~~\tilde{\mathcal{C}}=-\bm P_i \bm J_i\,. 
\label{Cas}
\ee
The possible quantum numbers $(k,\pm)$ for UIRs are 
\begin{align}
    \mathcal{C}A^{k\pm}_\mu& =-\frac{(k+1)^2}{\ell^2} A^{k\pm}_\mu,~~~~~ \tilde{\mathcal{C}} A^{k\pm}_\mu  = \mp \frac{(k+1)}{\ell } A^{k\pm}_\mu ,~~k=1,2,...
    \label{eigen}
\end{align}
Here each $k\in\mathbb N$ gives a different SO(4) UIR. Making $\bm k\mapsto\mathcal{L}_{\bm k}$ in \eqref{Cas} one finds (see \cite{AGS})
\begin{align}
\mathcal{C}A^{k\pm}_\mu=\left(\nabla^2-\frac2{\ell^2}\right)A^{k\pm}_\mu,~~~~~ \tilde{\mathcal{C}} A^{k\pm}_\mu= 
    \epsilon_\mu{}^{\alpha\beta} \nabla_\alpha A^{k\pm}_\beta  
    \label{lie}
\end{align}
here $\epsilon_{\mu\nu\rho}$ is the covariant Levi+Civita tensor. Comparing \eqref{eigen}-\eqref{lie} with the definition of $\Delta$ in \eqref{LV} we
have $\Delta=-k$ and $\Delta^s=2+k$. Since the lowest possible eigenvalue corresponds to $k=1$, we verify from \eqref{nkil} that the six Killing vectors  of the 3-sphere, spelled in \eqref{Js},\eqref{Ps}, are the lowest eigenvectors $(1,\pm)$ of the vector Laplacian. We can rewrite \eqref{ladderA} as
\begin{align}
{\cal D} A^{k\pm}_\mu&= c^\rho \nabla_\rho A^{k\pm}_\mu-k\, \sigma_{\bm c}\,  A^{k\pm}_\mu +\frac{1}{k}\nabla_\mu( c^\rho A^{k\pm}_\rho)\,. 
\end{align}
Analogously, for the shadow ladder  we have
\begin{align}
{\cal D}^s A^{k\pm}_\mu&=c^\rho \nabla_\rho A^{k\pm}_\mu+(2+k) \, \sigma_{\bm c} A^{k\pm}_\mu-\frac{1}{2+k} \nabla_\mu( c^\rho A_\rho^{k\pm} )
\end{align}
From \eqref{DAA} we get
\begin{align}
    [{\ell^2}\nabla^2, {\cal D}] A^{k\pm}_\mu&=(1+2k)\, {\cal D}  A^{k\pm}_\mu\nn \\
    [{\ell^2}\nabla^2, {\cal D}^s] A^{k\pm}_\sigma&=-( 3+2k)\, {\cal D}^s A^{k\pm}_\sigma\,,
\end{align}
which lead to 
\begin{align}
    {\cal C\,( D} \bm A^{k \pm})&= -\frac{k^2}{\ell^2} {\cal D}  \bm A^{k \pm}\nn\\
    {\cal C\,( D}^s \bm A^{k \pm})&= -\frac{(k+2)^2}{\ell^2} {\cal D}^s  \bm A^{k \pm} 
\end{align}
This means that ${\cal D}^s,\cal D$ rise/lower the $k$-quantum number (cf. \eqref{eigen})
\begin{align}
{\cal D}&:~~~k\mapsto k-1\nn\\
{\cal D}^s&:~~~k\mapsto k+1\nn
\end{align}
which can be equivalently phrased as  \eqref{shiftSc}. 

In addition we find that
\begin{align}
    [\tilde{\cal C} ,{\cal  D}] \bm A^{k \pm} &=\pm {\cal D} \bm A^{k \pm}  \nn\\
    [\tilde{\mathcal C} ,{ \cal  D}^s] \bm A^{k \pm} &=\mp {\cal D}^s \bm A^{k \pm} 
\end{align}
These transformations imply that 
\begin{align}
    \tilde{\mathcal{C}} \left({\cal D} \bm A^{k\pm} \right)=&\mp \frac{k}{\ell }\, {\cal D} \bm A^{k\pm}  \\
    \tilde{\mathcal{C}}\left( {\cal D}^s \bm A^{k\pm}\right)=&\mp \frac{k+2}{\ell }\, {\cal D}^s \bm A^{k \pm} .
\end{align}
Since no UIR with $k=0$ exists, it is reassuring to verify that 
$${\cal D} \bm A^{k\pm}|_{k=1}=0.$$
This follows from \eqref{KilAn} and the fact that the $(k=1,\pm)$ are the six Killing vectors of S$^3$.

\end{document}